\documentclass[amssymb,amsfonts,12pt,cite]{iopart}

\newcommand{\lb}[1]{\label{#1}}

\newcommand{\Lra}{\Leftrightarrow}
\newcommand{\bc}{\begin{center}}
\newcommand{\ec}{\end{center}}
\newcommand{\be}{\begin{equation}}
\newcommand{\ee}{\end{equation}}
\newcommand{\bea}{\begin{eqnarray}}
\newcommand{\eea}{\end{eqnarray}}
\newcommand{\ba}[1]{\begin{array}{#1}}
\newcommand{\ea}{\end{array}}

\newcommand{\bt}[1]{\begin{table}[ht]\centering\begin{tabular}{#1}}
\newcommand{\et}[1]{\end{tabular}\caption{\small#1}\end{table}}

\newcommand{\sign}{\,{\mathrm{sign}}\,}
\newcommand{\diag}{{\mathrm{diag}}}
\def\e{{\,\rm e}\,}
\renewcommand{\theequation}{\thesection.\arabic{equation}}

\begin{document}

\title[Rot. Electric Class. Sol. of $(2+1)$~D $U(1)$ Einstein Maxwell Chern-Simons]{Rotating Electric Classical Solutions of\\ $2+1$~D $U(1)$ Einstein Maxwell Chern-Simons}

\author{P. Castelo Ferreira}
\address{CENTRA, Instituto Superior T\'ecnico, Av. Rovisco Pais,\\ 1049-001 Lisboa, Portugal\\ and \\
Departamento de F\'{\i}sica, Universidade da Beira Interior,\\ Rua Marqu\^es D'\'{A}vila e Bolama, 6200-081 Covilh\~a}

\begin{abstract}
We study electric stationary radial symmetric classical solutions of the $U(1)$
Einstein Maxwell Chern-Simons theory coupled to a gravitational massless scalar field with a cosmological constant in $2+1$ dimensions.
Generic aspects of the theory are discussed at an introductory level.
We study solutions for both negative sign (standard) and positive sign (ghost) of the gauge sector
concluding that although the expressions for the solutions are the same, the constants as well as the physics
change significantly. A rotating electric point particle is found.
For the standard sign and specific values of the parameters corresponding to solutions with positive
mass the singularity is dressed (in the sense that itself constitutes an horizon). The space-time
curvatures can be both positive or negative depending on the dominance of the scalar or topologically massive matter.
The Chern-Simons term is responsible for interesting
features, besides only allowing for rotating solutions, it imposes restrictive bounds on the
cosmological constant $\Lambda$ such that it belongs to a positive interval and is switch on
and off by the topological mass $m^2$. Furthermore the charge, angular momentum and mass of the
particle solution are expressed uniquely as functions of the ratio between the cosmological constant
and the topological mass squared $x=\Lambda/m^2$. The main drawback of our particle solution is that
the mass is divergent. Our background is a rotating flat space without angular deficit. We briefly
discuss parity and time-inversion violation by the Chern-Simons term which is explicit in the solutions
obtained, their angular momentum only depends on the relative sign between the Chern-Simons
term and the Maxwell term. Trivial solutions are briefly studied holding non-singular extended configurations.
\end{abstract}

\vspace{2pc}
%\noindent{\it Keywords}: Chern-Simons, Classical Solutions, Scalar Field
%\vspace{2pc}
%\noindent{\it eprint}: hep-th/0506244

%\tableofcontents

\maketitle

\setcounter{equation}{0}
\section{Introduction\lb{sec.int}}

Several works have studied three dimensional classical gravitational configurations on
topological and non topological field theories. The first works addressed Einstein
theories, the well known AdS BTZ black hole~\cite{BTZ_01}, Einstein Maxwell Chern-Simons theory~\cite{Kogan1,Kogan2} and rotating BTZ~\cite{BTZ_02} (see also~\cite{Carlip1}).

This work studies the classical solutions for a $2+1$D
Einstein Maxwell Chern-Simons theory coupled to a gravitational
massless scalar field (that is often interpreted as a dilaton field in string-frame).
It therefore extends the work already done, both in
Einstein Maxwell Chern-Simons theories~\cite{CS_02,CS_03,CS_03a,CS_04,CS_05,CS_06},
Einstein Maxwell theories~\cite{BTZ_03,lemos_01}, Einstein Maxwell theories with
dilatonic potentials~\cite{CM_01,CM_02}
and the more recent Dilaton Einstein Maxwell theories~\cite{lemos_02,SLO_01,SLO_02,SLO_03,SLO_04,SLO_05}.

Here we exclusively address pure electric solutions of
$3D$~Einstein Maxwell Chern-Simons coupled to a massless scalar field. We try to present in
a pedagogical way both general results and details of calculations.
In particular our action resembles closely the one of the works~\cite{CM_01,CM_02} together with a Chern-Simons term.
However we start from a more generic action only particularizing the action due to the inexistence of other
possible solutions.

The motivation to study our enlarged theory is two folded:
the quantum consistence of the theory,
and the embedding of a $3D$ system in a $4D$ world.
First demanding quantum consistence of the theory we have to
consider the Maxwell-Chern-Simons theory. Neither the pure Maxwell theory,
neither the Chern-Simons theory are consistent at quantum level.
If we start just with a Maxwell action, radiative (quantum) corrections will
induce the Chern-Simons term and if we start with just a Chern-Simons action,
quantum corrections will induce a Maxwell term, this correction is exact to
all orders~\cite{qed3_01,qed3_02} (see also~\cite{dunne} for a review).
Secondly our world is $4D$, therefore by counting degrees of freedom we need a
gravitational scalar field in a $3D$ physical systems. Although several
ways to embed $2+1$ dimensional systems in $3+1$ dimensions, the existence of a gravitational massless scalar field
is rather well established. Considering a dimensional reduction scheme we obtain what is called
Dilaton~\cite{MSW,Polchinski}. Alternatively one can consider the gauging under some symmetry
that effectively reduces the dimensionality of the problem, this is the example of
the massless scalar field of the works on polarized cylindrical
gravitational waves in $3+1$ gravity~\cite{QG_01,QG_02,lemos_03,QG_03,QG_04}.
It is not clear that this scalar field can always be interpreted as a dilaton field although for some
particular actions it can be proved that it correspond to dilaton in string frame~\cite{CM_01,CM_02}.

We also note that most of the literature in Abelian gauge
Chern-Simons address (anti-)self-dual solutions. Here we address pure electric solutions.

The article is organized in the following way. In section~\ref{sec.gen}
we present and discuss generic results of the Einstein Maxwell Chern-Simons theory coupled to a scalar field.
First we introduce and justify the Action. From it we derive the equations of motion and choose
a suitable metric parameterization. Also we derive the charge, angular momentum and the mass in the ADM formalism.
In section~\ref{sec.E} we solve the equations of motion in the Cartan-frame. In section~\ref{sec.sing}
we compute the curvature, investigate the existence of singularities and horizons. Then
in section~\ref{sec.MJQ.f} we compute the charge, angular momentum and mass for the configurations
obtained. Finally in section~\ref{sec.conc} we summarize the solutions obtained and discuss them.
In appendix~\ref{A.cartan} we introduce the Cartan Frame formalism (also known as non-coordinate frame)
and derive the equations of motion and other useful formulae.

\setcounter{equation}{0}
\section{General Results\lb{sec.gen}}
\subsection{Action and EOM\lb{sec.EOM}}

We take a generic $2+1$D Einstein Gravity coupled to a massless scalar field with a
Gauge Sector described by $U(1)$ Maxwell-Chern-Simons
\be
\hspace{-15mm}
\ba{rl}
S=&\displaystyle\frac{1}{2\pi}\int_M d^3x\left\{\sqrt{-g}\left[e^{a\phi}\left(R+2\lambda(\partial\phi)^2\right)-e^{b\phi}\Lambda\right.\right.\\[5mm]
&\displaystyle\left.\left.+\hat{\epsilon}\frac{e^{c\phi}}{2}\left(F_{\mu\nu}F^{\mu\nu}+J^\mu A_\mu\right)\right]-\hat{\epsilon}\frac{m}{2}\,\epsilon^{\mu\nu\lambda}A_\mu F_{\nu\lambda}\right\}
\ea
\lb{S}
\ee
where $a$, $b$, $c$, $\lambda$ and the cosmological constant $\Lambda$ are numerical parameters of the theory.
$\hat{\epsilon}=\pm 1$ simply sets the relative sign between the gauge sector and the gravitational sector.

Varying this action in relation to the fields $A_\mu$, $g^{\mu\nu}$ and $\phi$ we obtain
the equations of motion, i.e. the Maxwell, Einstein and scalar field equations
\be
\hspace{-15mm}
\ba{rcl}
\displaystyle \partial_\alpha(\sqrt{-g}e^{c\phi}F^{\alpha\mu})+\frac{m}{2}\,\epsilon^{\mu\alpha\beta}F_{\alpha\beta}&=&\sqrt{-g}e^{c\phi}J^\mu\\[5mm]
\displaystyle G_{\mu\nu}-a\nabla_\mu\partial_\nu\phi+a\,g_{\mu\nu}\nabla^2\phi+(\lambda-a^2)\partial_\mu\phi\partial_\nu\phi& &\\[2mm]
\displaystyle -\left(\frac{\lambda}{2}-a^2\right)g_{\mu\nu}(\partial\phi)^2+\frac{1}{2}e^{(b-a)\phi}g_{\mu\nu}\Lambda&=&\displaystyle 2e^{(c-a)\phi}T_{\mu\nu}\\[5mm]
\displaystyle e^{a\phi}\left[2(2a^2-\lambda)\nabla^2\phi+2a(2a^2-\lambda)(\partial\phi)^2\right]+(3a-b)e^{b\phi}\Lambda&=&\displaystyle \hat{\epsilon}(a+c)e^{c\phi}F^2
\ea
\lb{EOM}
\ee
where the Einstein and Stress-Energy tensors are defined as
\be
\ba{rcl}
\displaystyle G_{\mu\nu}&=&\displaystyle R_{\mu\nu}-\frac{1}{2}g_{\mu\nu}R\\[5mm]
\displaystyle T_{\mu\nu}&=&\displaystyle \hat{\epsilon}\left(F_{\mu\alpha}F_{\nu}^{\ \alpha}-\frac{1}{4}g_{\mu\nu}F^2\right)
\ea
\lb{GT}
\ee
and the covariant derivative and Laplacian are as usual
\be
\ba{rcl}
\nabla_\mu\partial_\nu\phi&=&\partial_\mu\partial_\nu\phi-\Gamma^\alpha_{\ \mu\nu}\partial_\alpha\phi\\[5mm]
\nabla^2\phi&=&\partial_\alpha\partial^\alpha\phi+\Gamma^\alpha_{\ \alpha\beta}\partial^\beta\phi\\[5mm]
\ea
\ee

Note that the scalar field equations presented are obtained from the usual
variation of the action with respect to $\phi$
\be
e^{a\phi}\left[a\,R-2\lambda\nabla^2\phi-a\lambda(\partial\phi)^2\right]-be^{b\phi}\Lambda=\hat{\epsilon}\,c\,e^{c\phi}F^2
\ee
summed with the contraction of the 3 Einstein equations with the metric times $2a$. In this way the
gravitational curvature is not present in equation~(\ref{EOM}).

Our convention for the Ricci tensor is
\be
R_{\mu\nu}=-\Gamma^\alpha_{\ \mu\alpha,\nu}+\Gamma^\alpha_{\ \mu\nu,\alpha}-\Gamma^\alpha_{\ \mu\beta}\Gamma^\beta_{\ \nu\alpha}+\Gamma^\alpha_{\ \beta\alpha}\Gamma^\beta_{\ \mu\nu}
\label{Ricci}
\ee
we note that when considering a cosmological constant $\Lambda$ the symmetric
definition of the Ricci tensor, maintaining the same metric signature,
is not equivalent and will account for the opposite sign of $\Lambda$.
In order to justify this choice, in the next subsection, we give the example
of $3$-dimensional deSitter space, a known and well studied
example with $\Lambda>0$.

\subsection{Metric, Ricci Tensor and Maxwell Tensor\lb{sec.metric}}

We take several parameterizations of a radial symmetric metric, in polar
coordinates $x^0=t$, $x^1=r$ and $x^2=\varphi$ of the form
\be
ds^2=g_{tt}dt^2+dr^2+g_{\varphi\varphi}d\varphi^2+2g_{t\varphi}dtd\varphi
\lb{gpar_gen}
\ee
with signature $(-,+,+)$.

The Antisymmetric tensor has only the non vanishing components
\be
F_{tr}=E_*\ \ \ \ \ \ \ F_{r\varphi}=B_*
\lb{EBcov}
\ee

All the functions $g_{tt}$, $g_{\varphi\varphi}$, $g_{t\varphi}$, $E_*$, $B_*$ and $\phi$
are radial symmetric, i.e. are $r$ dependent only.

There is a couple of important well establish points to stress to fully justify this ansatz.

The motivation of introducing the $g_{t\varphi}$ component of the metric is due
to the Maxwell equations, in the presence of the Chern-Simons term (without external currents),
not allowing for solutions $B_*=0$ or $E_*=0$ when $g_{t\varphi}=0$~\cite{Kogan1}  (both must be null or both must be present).
So when there is a Chern-Simons term in the action and we are considering only Electric or only Magnetic fields,
we must have $g_{t\varphi}\neq 0$, otherwise both fields are null. In physical terms means that the space-time is rotating,
although it can still be stationary as long as $g_{t\varphi}$ does not depend on the time coordinate.

Also one may consider a non null $F_{t\varphi}$ but for the metric parameterizations considered here
the Maxwell Equation in~(\ref{EOM}) for $\mu=1$ imposes it to be null.

Finally it is important to stress that one can add a generic parameterization for $g_{rr}=1/L^2$
by introducing a new radial coordinate $\rho$ such that $d\rho/dr=L$. This accounts for
a choice of coordinates and therefore does not change the physical results presented here.

Although in $4D$ space-time the choice of metric (most positive or most negative diagonal)
is not relevant, in $3D$ space-time one needs extra care in the relative definitions
between the metric and remaining tensor fields. The reader may also note that depending on the
choice of $3D$ Minkowski metric the determinant is positive (for most negative diagonal)
or negative (for most positive diagonal). In~(\ref{gpar_gen}) we choose the last case to
maintain the determinant of the metric negative. To justify the choice of the Ricci
tensor~(\ref{Ricci}) and clear any confusions concerning its definition we present a
simple pedagogical example of the well known dS geometry which has positive cosmological constant.
We consider a cosmological Einstein action
\be
S_E=\int d^3x\sqrt{-g}\left(R-2\Lambda\right)
\ee
and a dS metric for an observer at r=0 corresponding to
a cosmological constant $\Lambda=+1$, of the form~\cite{ds}
\be
ds^2=-(1-r^2)dt^2+\frac{1}{(1-r^2)}dr^2+r^2d\varphi^2
\ee
with signature $(-,+,+)$ near the origin (where the observer is) and determinant $|g|=-r^2$.
Varying the action with respect to $g^{\mu\nu}$ we obtain the well know
equations of motion
\be
G_{\mu\nu}+\Lambda\,g_{\mu\nu}=0
\ee 
where $G_{\mu\nu}=R_{\mu\nu}-g_{\mu\nu}R/2$ is the usual Einstein tensor.
For the given metric, computing explicitly the einstein tensor, we obtain
$G_{00}=1-r^2$, $G_{11}=-1/(1-r^2)$ and $G_{22}=-r^2$. This reads
\be
G_{\mu\nu}=-g_{\mu\nu}
\ee
Therefore the cosmological constant is uniquely define trough the
equations of motion as $\Lambda=+1$. Maintaining the metric signature
and the action and considering the symmetric definition of the Ricci tensor
$\tilde{R}_{\mu\nu}=-R_{\mu\nu}$ we would obtain $\tilde{G}_{\mu\nu}=-G_{\mu\nu}$
and hence $\Lambda=-1$. Together with the definition $\tilde{R}$, if we swap the
signature of the metric to $(+,-,-)$ maintaining the action or if we maintain the
signature of the metric and change the action to $\tilde{S}_E=\int(R+2\Lambda)$
we would obtain $\Lambda=+1$. Also using the definition $R$, swapping the signature
of the metric to $(+,-,-)$ and considering the action $\tilde{S}_E$ we would
obtain $\Lambda=+1$.

So we conclude that the choices of the definition of the Ricci tensor,
the metric signature and the relative sign of the cosmological constant and the
gravitational curvature in the action are not all equivalent. Resuming, we choose
the definition of the Ricci tensor given by~(\ref{Ricci}), the metric signature $(-,+,+)$
and an action of the form~(\ref{S}).

Finally we briefly discuss the relative sign between the several terms in the action.
First we note that we consider opposite signs between the Chern-Simons term the Maxwell term.
This is to ensure that the photon mass is real
$(\nabla^2-m^2)F^*=0$~\cite{CS_grav_01,CS_grav_02}, if they have the same sign we would
obtain imaginary (tachyonic) masses. In particular this choice sets the sign of the angular
momentum $J$, as we will se our solutions have $J\sim m$ (or $\sign(m)$).
This is an effect of parity violation and is expected because the Chern-Simons
term violates parity in the gauge sector. If we change the relative sign between
the Maxwell term and the Chern-Simons term the only effect on the
solutions is to change the sign of the angular momentum. However as we explained
this accounts for the photon to become a tachyon, for this reason we fixed this choice.

$\hat{\epsilon}=\pm 1$ sets the relative sign between the gauge sector (Maxwell term $F^2$) and the
Einstein term ($R$). Choosing $\hat{\epsilon}=+1$ or $\hat{\epsilon}=-1$ does not
change the expressions for the solutions, nevertheless the
validity range for the parameters will change significantly, therefore
the physical interpretation of the results as well. Also it is interesting to note
that upon quantization the sign of the Maxwell is relevant. If we have
$\hat{\epsilon}=-1$ we obtain the standard Hamiltonian and
excited states of the gauge fields will have positive energy for Bose-Einstein spin-statistics,
while for $\hat{\epsilon}=+1$ the excited states for the gauge fields will only hold positive energy
for Fermi-Dirac spin-statistics. In this case the gauge fields have the \textit{wrong} spin-statistics
and for that reason are commonly called ghosts. It is quite interesting that different choice of
signs will also at classical level hold significant differences as we will see in detail.

\subsection{Mass, Charge and Angular Momentum\lb{sec.MJQ}}

We are going to use the ADM formalism~\cite{ADM} (see~\cite{gravitation}), so we rewrite
the line element using a generic parameterization\footnote{This metric parameterization is not unique but it accounts
for the most generic parameterization for a stationary radial symmetric $2+1D$ metric}
\be
ds^2=-f^2dt^2+dr^2+h^2(d\varphi+Adt)^2
\lb{gpar}
\ee
and considering the Hamiltonian form of the action
\be
S=-2\pi\Delta t\int dr\left[-f{\mathcal{H}}+A{\mathcal{H}}^\varphi+A_0{\mathcal{G}}\right]+S_{\mathcal{B}}
\ee
where $S_{\mathcal{B}}$ stands for boundary terms due to the integration by parts of the terms containing $f'$, $f''$,
$A'$ and $A_0'$
\be
\hspace{-15mm}
S_{\mathcal{B}}=\frac{1}{2\pi}\int_{\partial M}d^2x\,\left[f\,2e^{a\phi}(2a\,h\,\phi'+h')+A\,e^{a\phi}\Pi_G+\hat{\epsilon}A_0\left(\Pi_{EM}-\frac{m}{2}A_\varphi\right)\right]
\ee
and the Hamiltonian, Momentum and Gauss constraints are respectively
\be
\ba{rcl}
{\mathcal{H}}&=&\displaystyle-\frac{2\Pi_G^2}{h^3}e^{a\phi}-2a\left(h\phi'e^{a\phi}\right)'-2h''e^{a\phi}+2\lambda h(\phi')^2e^{a\phi}+\Lambda\,h\,e^{b\phi}\\[5mm]
             &+&\displaystyle \hat{\epsilon}\left(\frac{e^{-c\phi}}{h}\left(\Pi_{EM}+\frac{m}{2}\,A_\varphi\right)^2+h\,e^{c\phi}\,(A'_\varphi)^2\right)\\[7mm]
{\mathcal{H}}^\varphi&=&\displaystyle(\Pi_G\,e^{a\phi})'-\hat{\epsilon}\left(\Pi_{EM}+\frac{m}{2}\,A_\varphi\right)A'_\varphi\\[7mm]
{\mathcal{G}}&=&\displaystyle\hat{\epsilon}\left(\Pi_{EM}-\frac{m}{2}\,A_\varphi\right)'\ .
\ea
\ee
$\sqrt{-g}=h\,f$ and the induced $2D$ metric is simply $h_{ij}=\diag(1,h^2)$.
The prime ($'$) means the usual derivation ($\partial_r$) with respect to $r$.
We note that $\hat{\epsilon}$ in the gauss constraint is optional once
it is a constraint of the gauge sectors only.

For the rotating radially symmetric configurations considered in this work
(see subsections~\ref{sec.EOM} and~\ref{sec.metric}) the only non vanishing
gravitational canonical momenta  conjugate to $h_{ij}$
is $\pi_G^{r\varphi}$ (conjugate to $h_{r\varphi}$) such that
\be
\Pi_G={\mathrm{Tr}}(\pi_G)=(\pi_G)^r_{\ \varphi}=-\frac{h^3A'}{f}
\ee
and the only non vanishing gauge canonical momenta conjugate
to $A_i$ is $\pi_{EM}^{r}=\delta S/\delta (\partial_0A_r)$ (conjugate to $A_r$) such that
\be
\Pi_{EM}=\hat{\epsilon}\left({\mathcal{E}}-\frac{m}{2}\,A_\varphi\right)
\ee

The contravariant \textit{Electric} and \textit{Magnetic} densities are defined as~\cite{gravitation}
\be
\ba{rcl}
{\mathcal{E}}&=&\displaystyle h\,f\,e^{c\phi}F^{0r}=\hat{\epsilon}\left(\Pi_{EM}+\frac{m}{2}\,A_\varphi\right)\\[5mm]
{\mathcal{B}}&=&\displaystyle h\,f\,e^{c\phi}\epsilon^{r\varphi}F_{r\varphi}=\,h\,f\,e^{c\phi}\,A'_\varphi
\ea
\lb{EBcontrav}
\ee
For completeness we also note that the contravariant current densities are defined as
\be
{\mathcal{J}}^\mu=h\,f\,e^{c\phi}J^\mu
\lb{Jdens}
\ee

There is a couple of important points that should be stressed.
Generally, due to the rotation,
magnetic configurations generate a magnetic field and magnetic configurations generate an electric field.
However we will solve our equations in the Cartan frame such that for given fields $E$ and $B$ in the Cartan frame
we obtain ${\mathcal{E}}=h\,e^{c\phi}E$ and ${\mathcal{B}}=h^2\,f\,e^{c\phi}B$. Therefore we don't actually
have mixing between electric and magnetic fields (see appendix~\ref{A.cartan}). Ww also note that in $3D$ the
magnetic field is a scalar that corresponds in $4D$ to the $z$-component of the magnetic field, this means
the magnetic field perpendicular to the $2D$ spatial coordinates. In our configurations it is null.

The generic gauge canonical momenta are $\pi_{EM}^i=\hat{\epsilon}(hf\,e^{c\phi}F^{0i}-m\epsilon^{ij}A_j/2)$ and
therefore $\pi^\varphi_{EM}$ is not generally null.
However we are only studying configurations in which $F_{t\varphi}=\partial_tA_\varphi-\partial_\varphi A_t=0$
(see discussion on subsection~\ref{sec.metric}) such that $\pi_{EM}^\varphi=-m\,\hat{\epsilon}\,A_r$. Since we are considering only
rotating radial symmetric configurations we consider that all the gauge fields are radial functions,
furthermore we still have a radial gauge freedom, this means that a gauge transformation $\Lambda(r)$
depending on the radius only has the effect $A_r\to A_r+\Lambda'(r)$ and does not change any of the
physical quantities. Therefore we can without lost of generality gauge fix $\pi_{EM}^\varphi=A_r=0$.

As a final remark note that in the pure Maxwell theory ($m=0$) the canonical momentum is proportional to
the Electric density $\Pi_{\mathrm{Maxwell}}=\hat{\epsilon}{\mathcal{E}}$ such that this density is itself
a canonical variable, with the Chern-Simons term this is no longer true.

Varying both the action $S$ and the boundary action $A_{\mathcal{B}}$ with respect to the canonical
dynamical variables ($h,\Pi_G,\phi,\Pi_{EM},A_\varphi$) one obtains a boundary variation~\cite{BT}
\be
\delta S_{\mathcal{B}}=-2\pi\Delta t(-f\delta M+A_0\delta Q+A\delta J)
\ee
where $M$, $Q$ and $J$ are the Mass, Charge and Angular Momentum of the configuration and
$\mathcal{B}$ stands for the one-dimensional spatial boundary of the spatial manifold.
Their variation is
\be
\ba{rcl}
\delta M&=&\displaystyle\left.2\,\delta\left(h\,e^{a\phi}\right)'+4\lambda h\phi' e^{a\phi}\,\delta\phi+2\hat{\epsilon}h e^{c\phi}A'_\varphi\, \delta A_\varphi\right|_{\mathcal{B}}\\[5mm]
\delta Q&=&\displaystyle\left.2\hat{\epsilon}\,\delta\left(\Pi_{EM}-\frac{m}{2}\,A_\varphi\right)\right|_{\mathcal{B}}\\[5mm]
\delta J&=&\displaystyle\left.2\,\delta\left(\Pi_Ge^{a\phi}\right)-2\hat{\epsilon}\left(\Pi_{EM}+\frac{m}{2}A_\varphi\right)\,\delta A_\varphi\right|_{\mathcal{B}}
\ea
\lb{delMJQ}
\ee
In order to exist well defined classical minimum it is necessary that these variations vanish.
We need either to add a boundary action that cancels these variations or to demand them
(the variations) to vanish at the boundary. The later is usually a very strong condition
and accounts for having expressions for M, Q and J to be constants (meaning $r$ independent).
In the absence of external currents the charge $Q$ is necessarily
a constant since the Gauss' law is expressed as a total derivative. Accounting with the charge
expression, the angular momentum $J$ is also expressed as a total derivative and is therefore
a constant as well. For the case of the mass $M$ this is no longer true and we need to add a suitable
boundary action. In the presence of external currents neither $Q$ nor $J$ are generally constants since
the Gauss' law includes the external charge and is no longer a total derivative, here we are not addressing this
case.

Considering the above procedure we obtain
\be
\ba{rcl}
M&=&\displaystyle\left.2(h\,e^{a\phi})'+4\lambda h\phi\phi' e^{a\phi}+2\hat{\epsilon} h e^{c\phi}A_\varphi A'_\varphi\right|^{r\to\infty}_{r\to 0}\\[5mm]
Q&=&\displaystyle\left.2\hat{\epsilon}\left(\Pi_{EM}-\frac{m}{2}\,A_\varphi\right)\right|^{r\to\infty}_{r\to 0}\\[5mm]
J&=&\displaystyle\left.2\Pi_G e^{a\phi}-2\hat{\epsilon}\left(Q+\frac{m}{2}\,A_\varphi\right)A_\varphi\right|^{r\to\infty}_{r\to 0}
\ea
\lb{MJQ}
\ee
where we used the fact that
once the charge constraint in equation~(\ref{delMJQ}) is taken care, the charge variation
vanishes $\delta Q=0$, and used the expression for the charge to replace
$\Pi_{EM}=Q+m\,A_\varphi/2$ in the second term of the equation for the angular
momentum variation in order to get a variation of $A_\varphi$ only. We are considering
two disconnected boundaries, the spatial infinite $r=\infty$ and the singularity at
the origin $r=0$. We note that these two boundaries have opposite orientations, such that
their contributions add up.

As for the mass expression we have to be careful with what fields are fixed
and what fields vary upon a functional variation.
The correct expression should be
\be
M=\left.2(h\,e^{a\phi})'+4\lambda h\phi\hat{\phi}' e^{a\hat{\phi}}+2\hat{\epsilon} h e^{c\hat{\phi}}A_\varphi \hat{A}'_\varphi\right|^{r\to\infty}_{r\to 0}
\ee
where the hatted fields are fixed at the two boundaries ($r\to 0$ and
$r\to\infty$), i.e. upon a functional variation of the mass we obtain the
correct expression~(\ref{delMJQ}).

\subsection{Geodesics and Horizons}

To compute the geodesics we use the variational principle presented in~\cite{ray}, so we consider
the constant functional
\be
K=g_{\mu\nu}\frac{x^\mu}{ds}\frac{x^\nu}{ds}=\kappa=
\left\{\ba{rl}
0&\mathrm{for\ lightlike\ (null)\ geodesics}\\
-1&\mathrm{for\ timelike\ geodesics}\\
+1&\mathrm{for\ spacelike\ geodesics}
\ea\right.
\lb{K_E}
\ee
where the derivatives are with respect to a affine parameter $s$. We minimize $K$
solving the Euler-Lagrange equations $\frac{\delta K}{\delta x^\mu}-\frac{d}{ds}\left(\frac{\delta K}{\delta\dot{x}^\mu}\right)=0$.
Since our solutions are both cylindrically symmetric and stationary (only depend on $r$, the radial coordinate)
we have that the equations for $\mu=t,\varphi$ lead respectively to the first integrals of motion
\be
\left\{\ba{rcl}
\displaystyle g_{00}\,\frac{dt}{ds}+g_{02}\,\frac{d\varphi}{ds}&=&\displaystyle E\\[5mm]
\displaystyle g_{22}\,\frac{d\varphi}{ds}+g_{02}\,\frac{dt}{ds}&=&\displaystyle L
\ea\right.
\Rightarrow
\left\{\ba{rcl}
\displaystyle \frac{dt}{ds}&=&\displaystyle\frac{E\,g_{22}-L\,g_{02}}{g}\\[5mm]
\displaystyle \frac{d\varphi}{ds}&=&\displaystyle\frac{L\,g_{00}-E\,g_{02}}{g}
\ea\right.
\lb{dtdfEL_E}
\ee
with $2E=p_t=\frac{\delta K}{\delta\dot{t}}$ and $2L=p_\varphi=\frac{\delta K}{\delta\dot{\varphi}}$
being constants of motion, the energy and angular momentum respectively (here we rescaled them by a factor of $2$
in order to simplify the expressions).
Using the two equations~(\ref{dtdfEL_E}) in~(\ref{K_E}) we obtain an expression for $dr/ds$
\be
\left(\frac{dr}{ds}\right)^2=k-\frac{L^2\,g_{00}-2E\,L\,g_{02}+E^2\,g_{22}}{g}
\ee
being $g$ the determinant of the metric $g=g_{00}g_{22}-g_{02}^2$.

Since we are looking for stationary polar symmetric solutions $\dot{t}$ and $\dot{\varphi}$
can be expressed in terms of the radial variable $r$ only, $(d/ds)/(dr/ds)=d/dr$.
From the equations for $t$ and $\varphi$~(\ref{dtdfEL_E}) we obtain the differential equations
\be
\ba{rcl}
t'(r)&=&\displaystyle\pm\frac{E\,g_{22}-L\,g_{02}}{\sqrt{g\left(g\kappa-L^2\,g_{00}+2E\,L\,g_{02}-E^2\,g_{22}\right)}}\\[5mm]
\varphi'(r)&=&\displaystyle\pm\frac{L\,g_{00}-E\,g_{02}}{\sqrt{g\left(g\kappa-L^2\,g_{00}+2E\,L\,g_{02}-E^2\,g_{22}\right)}}
\ea
\lb{dtdf_dr}
\ee
Solving these equation one obtains the $t$ and $\varphi$ dependence on $r$.
We can also compute the radial velocity $\dot{r}=(dr/ds)/(dt/ds)$ and angular velocity $\dot{\varphi}=(d\varphi/ds)/(dt/ds)$
\be
\ba{rcl}
\dot{r}(r)&=&\displaystyle\pm\frac{\sqrt{g\left(g\kappa-L^2\,g_{00}+2E\,L\,g_{02}-E^2\,g_{22}\right)}}{E\,g_{22}-L\,g_{02}}\\[5mm]
\dot{\varphi}(r)&=&\displaystyle\frac{L\,g_{00}-E\,g_{02}}{E\,g_{22}-L\,g_{02}}
\ea
\lb{drdf_dt}
\ee

We note that these solutions are for an external observer (at rest far away from the singularity).
Then the first equation is particular useful, when $\dot{r}=0$ we are in the presence either of a turning point
on the trajectory, or of a horizon (in which case the geodesics at the rest frame of the travelling observer
hits the singularity). We also note that at the singularity, if $\dot{r}$ is null the singularity is
not naked, meaning that an external observer sees the particle stopping when arriving to the singularity.
While if $\dot{r}$ has some positive value at the singularity we have a naked singularity since an external
observer can actually see it without reaching it. We are using this results to inquire if we have an horizon or not.

\setcounter{equation}{0}
\section{Electric Solutions\lb{sec.E}}

Here we will look for pure Electric solution without external currents, hence we set $B=B_*$,
being $B$ the magnetic field in the Cartan frame and $B_*$ the magnetic field in the original frame.
We will be working in the Cartan frame and at the end of each subsection we will summarize
our results in the original frame. The equations of motion in the Cartan frame are computed
in appendix~\ref{A.cartan} and are equivalent to the equations of motion as presented in subsection~\ref{sec.EOM}.

From the first Maxwell Equation~(\ref{E.M1}) we obtain that
\be
\gamma=m\,e^{-c\phi}
\lb{E.gam}
\ee

Using~(\ref{E.gam}) in~(\ref{E.E1}) one gets that $\beta=c \phi'/2$ and from the definition of $\beta$
(see~(\ref{E.abgdef}) in appendix) we get the solution for $h$
\be
h=c_h\,e^{\frac{c}{2}\phi}
\lb{E.ha}
\ee
where $c_h$ is a free integration constant.

Now we get from the second Maxwell Equation~(\ref{E.M2}) that
\be
E=\chi e^{-\frac{3}{2}c\phi}
\lb{E.E}
\ee
where $\chi$ is an integration constant. Note that without loss of generality we included $c_h$
in the definition of this constant. There is a very important conclusion to take from this last
equation, trivial solutions for the scalar field ($\phi=\mathrm{constant}$) holds in the Cartan frame a uniform (constant)
electric field $E$ in all space, this conclusion was firstly obtained in~\cite{Kogan1}. 
Although for completeness we address trivial solutions
we will first address non-trivial solutions for the scalar field which is the main objective of this work.

\subsection{Non-Trivial Solutions for the Scalar Field}

We will now address the full equations considering the generic equations.
The three Einstein~(\ref{E.E2}-\ref{E.E4}) and scalar field equations~(\ref{E.D}) read now
\bea
\hspace{-19mm}\displaystyle (a+\frac{c}{2})\phi''+(a^2-\frac{\lambda}{2}+\frac{c^2}{4})(\phi')^2+\frac{m^2}{4}e^{-2c\phi}+\frac{\Lambda}{2} e^{(b-a)\phi}&=&\displaystyle -\hat{\epsilon}\chi^2e^{(-a-2c)\phi}\lb{E.b2}\\[5mm]
\hspace{-19mm}\displaystyle a\phi''+(a^2-\frac{\lambda}{2})(\phi')^2+\alpha^2+\alpha'-\frac{3m^2}{4}e^{-2c\phi}+\frac{\Lambda}{2} e^{(b-a)\phi}&=&\displaystyle \hat{\epsilon}\chi^2e^{(-a-2c)\phi}\lb{E.b3}\\[5mm]
\hspace{-19mm}\displaystyle \frac{\lambda}{2}(\phi')^2+\frac{c}{2}\alpha\phi'+\frac{m^2}{4}e^{-2c\phi}+\frac{\Lambda}{2} e^{(b-a)\phi}&=&\displaystyle -\hat{\epsilon}\chi^2e^{(-a-2c)\phi}\lb{E.b4}\\[5mm]
\hspace{-19mm}\displaystyle (4a^2-\lambda)\phi''+a(4a^2-2\lambda)(\phi')^2+(3a-b)\Lambda e^{(b-a)\phi}&=&\displaystyle -\hat{\epsilon}(a+c)\chi^2e^{(-a-2c)\phi}\lb{E.b5}
\eea
The main problem to solve these equations is to make them compatible with each other in order to give
a non trivial solution.  For $a=b=c$, for $a=0$ (any $b$ and $c$), for $b=0$ (any $a$ and $c$) and $c=0$ (any $a$ and $b$)
these equations hold that the scalar field has only trivial solutions, i.e. it must be a constant.
Trivial solutions will be addressed in the next subsections. For the particular cases $c=0$ with
$a=b$ and $a=b=-2c$ solutions do exist but hold that the scalar field is purely imaginary.

The better way to properly understood the structure of the equations is the following. The third equation~(\ref{E.b4})
can be algebraically solved in $\alpha$ which solution is then plugged into the second equation~(\ref{E.b3}). Then
to obtain a solution for the $\phi$ we can make a linear combination of the remaining three equations
obtaining a simpler equation. The main problem then is to ensure that the solution is compatible
with the original equations (or equivalently with different linear combinations of the original equations).
This procedure gives very few choices for non-trivial solutions.

We only found non-trivial solutions for the case
\be
\ba{rcl}
a&=&0\\[5mm]
c&=&\displaystyle -\frac{b}{2}\\[5mm]
\lambda&\neq&\displaystyle \frac{b^2}{8}
\ea
\lb{choice_abc}
\ee
For $b^2=8\lambda$ does not exist a non-trivial solution either.
We note that for the choice of equation~(\ref{choice_abc}) we are not working with dilaton Einstein theory. Our action is more similar
to what is commonly know as a gravitational scalar field~\cite{QG_01,QG_02,QG_03,QG_04} and the cosmological constant term resembles
a Dilaton potential~\cite{dilpot}~\footnote{Thanks to Dmitri Gal'tsov for this remark.}

Given this ansatz we combine~(\ref{E.b2}) with~(\ref{E.b5}) obtaining
\be
\phi'=\pm\sqrt{c_1}e^{b\,\phi}
\lb{E.dphi}
\ee
such that
\be
\phi=-\frac{2}{b}\ln(c_\phi(r-r_0))
\lb{E.dil}
\ee
Here
\be
c_\phi=\frac{|b|}{2}\sqrt{c_1}\ \ \ \ c_1=-2\frac{b^2(\hat{\epsilon}\chi^2+2\Lambda)+2\lambda(4\hat{\epsilon}\chi^2+2\Lambda+m^2)}{\lambda(b^2-8\lambda)}
\lb{E.C1a}
\ee
and without loss of generality we set the integration constant $r_0=0$
since it represents only a shift in the radial coordinate and all the solutions depend on the $\phi$ exponentials.
Note that the choice of sign in~(\ref{E.dphi}) depends on the sign of $b$ such that
in~(\ref{E.dil}) the argument of the logarithm is positive. Also we have to ensure
that $c_1$ is positive defined. Before doing so we use the $\phi$ solution~(\ref{E.dil})
in~(\ref{E.b2}). In order the equation to be solved we have to impose
\be
\chi^2=-\hat{\epsilon}\frac{2\Lambda(b^2+12\lambda)+4\lambda m^2}{b^2+24\lambda}
\lb{E.chi}
\ee
Now $c_1$ becomes
\be
c_1=4\,\frac{m^2-6\Lambda}{b^2+24\lambda}
\lb{E.C1}
\ee

From~(\ref{E.b4}) and the definition $\gamma=A'h/f$~(\ref{E.abgdef}) we get that
\be
\alpha=-\left(16\frac{\lambda}{b^2}+1\right)\frac{1}{2\,r}
\ee

Therefore from the definition of $\alpha=f'/f$~(see~(\ref{E.abgdef}) in the appendix) we obtain the solution for $f$
\be
f=c_f\,r^{-\frac{8\lambda}{b^2}-\frac{1}{2}}
\lb{E.f}
\ee
from~(\ref{E.ha}) we obtain the solution for $h$
\be
h=c_h\,\sqrt{r}
\lb{E.h}
\ee
and from~(\ref{E.gam}) we get the solution for $A$
\be
A=c_A r^{-\frac{8\lambda}{b^2}-1}+c_{A_\infty}
\lb{E.A}
\ee
where
\be
c_A=\frac{m\,c_f}{c_h\left(-\frac{8\lambda}{b^2}-1\right)}\sqrt\frac{1+\frac{24\lambda}{b^2}}{m^2-6\Lambda}
\ee
$c_f$, $c_h$ and $c_{A_\infty}$ are free constants.

Replacing these solutions in the remaining equation~(\ref{E.b3}) and demanding it to be obeyed we get that
\be
\lambda_\pm=\frac{b^2}{8}\,\frac{3\Lambda\mp\sqrt{\Lambda(2m^2-3\Lambda)}}{m^2-6\Lambda}
\lb{E.l}
\ee
We have to ensure that all these relations are possible and that do not correspond to
trivial solutions, in particular that $\chi^2>0$ and $C_1>0$.
Therefore for each $\hat{\epsilon}=\pm 1$ we have to choose the solution $\lambda_{\hat{\epsilon}}$ getting
\be
\ba{rcl}
\chi^2&=&\displaystyle\frac{1}{2}\left[-\hat{\epsilon}\Lambda+\sqrt{\Lambda(2m^2-3\Lambda)}\right]\\[5mm]
C_1&=&\displaystyle \frac{4}{b^2}\left[3\Lambda+m^2+3\hat{\epsilon}\sqrt{\Lambda(2m^2-3\Lambda)}\right]
\ea
\ee
Demanding positiveness of these expressions hold, independently of
$\hat{\epsilon}$ the same constraint on the cosmological
constant $\Lambda$ and topological mass $m$
\be
0<\Lambda<\frac{m^2}{2}
\lb{E.Lm2}
\ee

For the particular value of $\Lambda=m^2/6$ some of the expressions previously computed are not well defined.
It is necessary to rederive the solution using the same method. For $\hat{\epsilon}=+1$ we obtain
\be
\Lambda=m^2/6\ \ \ \ C_1=\frac{12m^2}{b^2}\ \ \ \ \chi^2=\frac{m^2}{6}\ \ \ \ \lambda=-\frac{b^2}{24}\ \ \ \ C_A=\frac{\sqrt{3}c_f}{c_h}\ .
\ee
All the other solutions remain the same up to replacement of the above constants.
For $\hat{\epsilon}=-1$ there are no allowed solutions at $\Lambda=m^2/6$.

For convenience we define the parameter $p$ which depends only on the ratio $\Lambda/m^2$
\be
p=-8\frac{\lambda}{b^2}=\frac{-3x+\hat{\epsilon}\sqrt{x(2-3x)}}{1-6x}\ \ \ \ \ \ \ \ \ x=\frac{\Lambda}{m^2}
\lb{E.p}
\ee

For clarity we summarize and rewrite the solutions computed above in the original frame,
\be
\ba{rcl}
\phi&=&\displaystyle -\frac{2}{b}\ln(C_\phi\, r)\\[7mm]
h&=&\displaystyle C_h\, \sqrt{r}\\[7mm]
f&=&\displaystyle C_f\,r^{p-\frac{1}{2}}\\[7mm]
A&=&\displaystyle C_A\,r^{p-1}+\theta\\[7mm]
E_*&=&\displaystyle C_E\,r^{p-2}\\[7mm]
A_0&=&\displaystyle\frac{C_E}{p-1}\,r^{p-1}
\ea
\lb{solsE}
\ee
where for convenience we rename the variables and integration constants.
$C_h$, $C_f$, $b$ and $\theta $ which are free parameters while the remaining variables are
\be
\ba{rcl}
p&=&\displaystyle -\frac{3\Lambda-\hat{\epsilon}\sqrt{\Lambda(2m^2-3\Lambda)}}{m^2-6\Lambda}\\[7mm]
\lambda&=&\displaystyle-\frac{8}{b^2}p\\[7mm]
C_\phi&=&\displaystyle \sqrt{\frac{m^2-6\Lambda}{1-3p}}\\[7mm]
C_A&=&\displaystyle\frac{m\,C_f}{C_h\left(p-1\right)}\sqrt\frac{1-3p}{m^2-6\Lambda}\\[7mm]
C_{E(\pm)}&=&\displaystyle\mp \frac{C_f}{\sqrt{2}}\sqrt{4\Lambda-p(m^2+6\Lambda)}\left(\frac{1-3p}{(m^2-6\Lambda)^3}\right)^\frac{1}{4}
\ea
\lb{constantsE}
\ee
Here $\theta=C_{A_\infty}$ in~(\ref{E.A}). For $\hat{\epsilon}=+1$ and the particular case $\Lambda=m^2/6$
corresponding to $p=1/3\Lra \lambda=-b^2/24$ we have $C_A=\sqrt{3}C_f/(2C_h)$. The values of the remaining constants
are well defined, $C_\phi=\sqrt{3}\,m$ and $C_{E(\pm)}=\mp 3\,C_f\,m^{5/2}/\sqrt{2}$.
For the values $p=0$ ($\Lambda=0$) and $p=1/2$ ($\Lambda=m^2/2$) we obtain $C_E=0$ and
therefore the solutions presented here do not allow charged configurations for these
particular limit values. In these cases $C_\phi\sim m$.
For $\hat{\epsilon}=-1$ the particular case $\Lambda=m^2/6$ has no real solutions.

We have the bound in the cosmological constant
\be
0<\Lambda<\frac{m^2}{2}
\ee
such that $p$ is in the range
\be
\left\{\ba{rclcl}
p&\in&\displaystyle\left]0,\frac{1}{2}\right[&\ \ \ \ &\hat{\epsilon}=+1\\[5mm]
p&\in&\left]-\infty,0\right[\cup\left]1,+\infty\right[&\ \ \ \ &\hat{\epsilon}=-1
\ea\right.
\ee
where for both cases $p=0$ corresponds to $\Lambda=0$ and for $\hat{\epsilon}=+1$ we have
$p=1/2$ corresponding to $x=\Lambda/m^2=1/2$ while for $\hat{\epsilon}=-1$ we have $p=1$ corresponding to $x=\Lambda/m^2=1/2$.
For $\hat{\epsilon}=+1$ we have that $p=1/3$ corresponds to $x=\Lambda/m^2=1/6$ while
for $\hat{\epsilon}=-1$ we have that $\lim_{x\to (1/6)^\pm}p=\mp\infty$. For $\hat{\epsilon}=-1$, $p\in]-\infty,0[$
corresponds to $x=\Lambda/m^2\in]0,1/6[$ and $p\in]1,\infty[$ corresponds to $x=\Lambda/m^2\in]1/6,1/2[$.

\subsection{Trivial Scalar Field Solutions: $\phi=0$}

It remains to analyse the case of $\phi=0$. This case corresponds to not considering the scalar field at all
and has been first addressed by Kogan~\cite{Kogan1}, however in the original work a cosmological constant
have not been considered (it has in~\cite{Kogan2} but without solving the equations of motion),
for this reason we also discuss it here.

Considering the above solutions for $\gamma$~(\ref{E.gam}), $h$~(\ref{E.ha}) and $E$~(\ref{E.E})
the remaining three Einstein~(\ref{E.E2}-\ref{E.E4}) reduce only to two independent equations
\bea
\displaystyle \frac{m^2}{4}+\frac{\Lambda}{2}&=&\displaystyle -\hat{\epsilon}\chi^2\lb{E.ttb1}\\[5mm]
\displaystyle \alpha^2+\alpha'-\frac{3m^2}{4}+\frac{\Lambda}{2}&=&\displaystyle \hat{\epsilon}\chi^2\lb{E.ttb2}
\eea
while the scalar field equation~(\ref{E.D}) is already obeyed.
Solving the first equation for $\chi^2$ we get
\be
\chi^2=-\hat{\epsilon}\left(\frac{m^2}{4}+\frac{\Lambda}{2}\right)
\lb{E.ttchi}
\ee
and demanding the right hand side to be positive definite we obtain the constraint
\be
\left\{\ba{rclcl}
\Lambda&<&-\frac{m^2}{2}&\ \ \ \ &\hat{\epsilon}=+1\\[5mm]
\Lambda&>&-\frac{m^2}{2}&\ \ \ \ &\hat{\epsilon}=-1
\ea\right.
\lb{L.ttbound}
\ee
As in the previous subsection in order to exist electric solutions the cosmological constant is
constraint to be negative for $\hat{\epsilon}=+1$ and can be both negative in the range $]-m^2/2,0[$
or positive for $\hat{\epsilon}=-1$. We also note that from~(\ref{E.ttchi}) the equality $\Lambda=m^2/2$
holds that $\chi=E=0$, therefore not allowing electric configurations. For
this reason we don't consider the case $\Lambda=-m^2/2$.

From equation~(\ref{E.ttb2}) and the definition of $\alpha$ (see~(\ref{E.abgdef}) in appendix) we get the solution
for $f$
\be
f=c_f\cosh\left(\sqrt{k}\,(r-r_0)\right)
\ee
where $c_f$ and $r_0$ are integration constants and
\be
k=\frac{m^2}{2}-\Lambda\ .
\ee
From~(\ref{E.ha}) we have that
\be
h=c_h
\ee
and from~(\ref{E.gam}) and the definition for $\gamma$~(\ref{E.abgdef}) we obtain the
solution for $A$
\be
A=\frac{m\,c_f\sinh\left(\sqrt{k}\,(r-r_0)\right)}{c_h\sqrt{k}}+c_{A_0}
\ee
where $c_{A_0}$ is an integration constant that corresponds to the value of $A$ at $r=r_0$.
Again we can set $r_0=0$ since it represents a shift in the radial coordinate.

For clarity we summarize and rewrite the solutions just obtained in the original frame
\be
\ba{rcl}
h&=&\displaystyle C_h\\[7mm]
f&=&\displaystyle C_f\,\cosh(K\,r)\\[7mm]
A&=&\displaystyle C_A\,\sinh(K\,r)+\theta\\[7mm]
E_*&=&\displaystyle C_E\,\cosh(K\,r)\\[7mm]
A_0&=&\displaystyle \frac{C_E}{K}\,\sinh(K\,r)
\ea
\lb{tsolsE}
\ee
where $C_h$ and $C_f$ are free constants and the remaining constants are defined as
\be
\ba{rcl}
K&=&\displaystyle\sqrt{\frac{m^2}{2}-\Lambda}\\[7mm]
C_A&=&\displaystyle\frac{m\,C_f}{C_h\, K}\\[7mm]
C_{E(\pm)}&=&\displaystyle\pm \frac{C_f}{2}\sqrt{\left|\frac{m^2}{2}+\Lambda\right|}
\ea
\lb{tconstantsE}
\ee
The cosmological constant is constraint and accordingly $K$ is real for $\hat{\epsilon}=+1$
\be
\hat{\epsilon}=+1\ :\ \ \ 
\left\{\ba{rcl}
\Lambda&<&\displaystyle-\frac{m^2}{2}\\[5mm]
K&\in&]m^2,+\infty[
\ea\right.
\ee
but can be both real and imaginary for $\hat{\epsilon}=-1$
\be
\hat{\epsilon}=-1\ :\ \ \ 
\left\{\ba{rcl}
\Lambda&\in&\displaystyle\left]-\frac{m^2}{2},0\right[\\[5mm]
K&\in&]0,m^2[
\ea\right.
\ \ \ \mathrm{or}\ \ \ 
\left\{\ba{rcl}
\Lambda&\in&\displaystyle\left]0.+\infty\right[\\[5mm]
K&\in&]0,+\infty[\ i
\ea\right.
\ee
where the last interval for $K$ is imaginary. In this last case we obtain periodic solutions in $r$ with period $2\pi/|K|$.

As a final remark we note that the contravariant electric density as defined in~(\ref{EBcontrav}) is a constant
\be
{\mathcal{E}}=-\frac{C_E\,C_h}{C_f}
\ee 
as expected from the solution for $E$ in the Cartan frame.

We already analyse the case for a null scalar field, but a constant scalar field is also an allowed
trivial solution. For such solutions we obtain the same solutions up to the redefinition of the
parameters
\be
\tilde{\Lambda}=\Lambda\,e^{(b-a)\phi}\ \ \ \ \ \tilde{m}=m\,e^{-c\phi}\ \ \ \ \ \tilde{\chi}=\chi\,e^{-\frac{a}{2}-c}
\ee 
with $\phi=\mathrm{constant}$.

\setcounter{equation}{0}
\section{Singularities, Geodesics and Horizons\lb{sec.sing}}

\subsection{Non-Trivial Solutions}

The contraction of the Ricci tensor is
\be
\hspace{-15mm}
\ba{rcl}
R_{\mu\nu}R^{\mu\nu}&=&\displaystyle\frac{1}{4C_f^4 r^4}[C_f^4(3+2p(-8+p(17+2p(-7+2p))))-\\[5mm]
                    & &\displaystyle 2(C_AC_fC_h(p-1))^2(3+4p(p-2))r+3(C_AC_h(p-1))^4]
\ea
\ee
which shows that there is a curvature singularity at $r=0$.
The curvature is
\be
\hspace{-15mm}
R=\frac{h^3A'^2-4f(hf''+f'h'+fh'')}{2f^2h}=\frac{m^2p(3-4p)+6\Lambda(4p^2-6p+1)}{2(m^2-6\Lambda)}\,\frac{1}{r^2}\ .
\ee
For $\hat{\epsilon}=+1$ we have always positive curvature while
for $\hat{\epsilon}=-1$ we can have both negative and positive curvatures
\be
\ba{rclrl}
r^2R&\in&\displaystyle\left]0,\frac{5}{8}\right[&\ \ \ \ \mathrm{for}&\displaystyle\hat{\epsilon}=+1\ \ \mathrm{and}\ \ x\in\left]0,\frac{1}{2}\right[\\[5mm]
r^2R&\in&\displaystyle\left]-\infty,0\right[&\ \ \ \ \mathrm{for}&\displaystyle\hat{\epsilon}=-1\ \ \mathrm{and}\ \ x\in\left]0,\frac{9}{38}\right[/\left\{\frac{1}{6}\right\}\\[5mm]
r^2R&=&0&\ \ \ \ \mathrm{for}&\displaystyle\hat{\epsilon}=-1\ \ \mathrm{and}\ \ x=\frac{9}{38}\\[5mm]
r^2R&\in&\displaystyle\left]0,\frac{9}{8}\right]&\ \ \ \ \mathrm{for}&\displaystyle\hat{\epsilon}=-1\ \ \mathrm{and}\ \ x\in\left]\frac{9}{38},\frac{1}{2}\right[
\ea
\ee
For the limiting cases $\Lambda\to 0$ (corresponding to $x\to 0$) we have $R\to 0$ for both $\hat{\epsilon}=\pm 1$ and
for $\Lambda\to m^2/2$ (corresponding to $x\to 1/2$) we have $R\to 5/(8r^2)$ for $\hat{\epsilon}=+1$
and $R\to 1/r^2$ for $\hat{\epsilon}=-1$. The Maximum value of the curvature for the case
$\hat{\epsilon}=-1$ is $R=9/(8r^2)$ corresponding to $x=9/26$. We note that as already explained in
the last section $x=0$ and $x=1/2$ are not allowed solutions and, for $\hat{\epsilon}=-1$, $x=1/6$
is neither an allowed solution.

For both cases $\hat{\epsilon}=\pm 1$ the curvature is asymptotically flat ($\lim_{r\to \infty} R=0$),
therefore our spaces are asymptotically flat.

In order to find if there is or not an horizon it is enough to consider a photon travelling in the radial direction.
So we can solve equations~(\ref{dtdf_dr}) with $L=0$ and $\kappa=0$ obtaining
\be
\ba{rcl}
t(r)&=&\displaystyle t_0\pm\frac{2}{|C_f|(2p-3)}r^{3/2-p}\\[5mm]
\varphi(r)&=&\displaystyle\varphi_0\pm\frac{2\left((2p-3)r^{p-1}-\theta\right)}{|C_f|(2p-3)}r^{3/2-p}
\ea
\lb{tphi_sing}
\ee
For $\hat{\epsilon}=+1$ we have that $p\in]0,1/2[$, so these solutions are regular for all $r$ and
we conclude that there is no horizon. From regularity at the singularity $r=0$
we are in the presence of a naked singularity, for an external observer the photon
will hit the singularity in a finite time.
For $\hat{\epsilon}=-1$ we can have an horizon at $r=0$ as long as $p>3/2$ ($p=3/2\Leftrightarrow x=\Lambda/m^2=9/26$).
This will happen for
\be
x=\frac{\Lambda}{m^2}\in\left]\frac{1}{6},\frac{9}{26}\right[\ .
\ee
Then in this range we will have a dressed singularity, for an external observer the infalling particle
will take an infinite amount of time to reach the singularity. For all other values of $p$
we have a naked singularity. We note that from~(\ref{tphi_sing}) for $p=3/2$ the geodesics are a fixed
point on time and there are no horizons. 

In order to understand the meaning of our singularity in terms of the angular variable
let us now compute the angle deficit of our space, or equivalently the maximum value for the
angular variable $\varphi$. The metric reads
\be
ds^2=-r^{2p-1}\,dt^2+dr^2+C_h^2 r\,\left(d\varphi+A\,dt\right)^2\ .
\ee
Let us remember from the discussion in section~\ref{sec.gen} that the $2D$~induced
metric is $h_{ij}=\diag(1,h^2)=\diag(1,C_h^2\,r)$. Now let us make a transformation
of coordinates $r\to \tilde{r}$ such that the measure of the induced metric is the usual one,
i.e $\sqrt{|h_{ij}|}=\tilde{r}^2$. This accounts for a observer at rest in relation
to space-time (hence rotating with space). The transformation of the radial coordinate is
\be
r=\left(\frac{3}{4C_h}\right)^\frac{2}{3}\tilde{r}^\frac{4}{3}\ \ \ \ \Rightarrow
\ \ \left\{\ba{rcl}f&=&\displaystyle\left(\frac{3\tilde{r}^2}{4C_h}\right)^{\frac{2}{3}\left(p-\frac{1}{2}\right)}\\[5mm]
                   h_{rr}&=&\displaystyle\left(\frac{4\tilde{r}}{3C_h^2}\right)^\frac{2}{3}\\[5mm]
                   h_{\varphi\varphi}&=&\displaystyle\left(\frac{3C_h^2\tilde{r}^2}{4}\right)^\frac{2}{3}\ea\right.
\ee
The maximum angle is computed as
\be
\varphi_{\mathrm{max}}=\frac{2\pi}{\sqrt{-g}}\sqrt{\frac{h_{\varphi\varphi}}{h_{rr}}}=\frac{2\pi}{f\,h_{rr}}\ .
\ee
In order to obtain the background geometry we take the limit
$p\to 0$ (equivalent to $\Lambda\to 0$). We will discuss this
limit properly in the next section when computing the mass, charge and angular momentum,
for now let us just take it as granted, then the respective maximum angle is
\be
\varphi_{\mathrm{max}}=2\pi\,\frac{3|C_h|}{4}\ .
\ee
Imposing it to be as usal $2\pi$ we obtain the value for $C_h$
\be
|C_h|=\frac{4}{3}\ .
\ee
So we have a rotating background without any angle deficit. For generic $p$ we obtain
\be
\varphi_{\mathrm{max}}=2\pi\,\left(\frac{3\tilde{r}}{4}\right)^{-\frac{4}{3}p}
\ee
such that for $\tilde{r}=4/3$ we have $\varphi_{\mathrm{max}}=2\pi$ always.
In the limit $\tilde{r}\to 0$ we obtain that for $p>0$, $\varphi_{\mathrm{max}}\to\infty$ and for $p<0$,
$\varphi_{\mathrm{max}}\to 0$. While in the limit $\tilde{r}\to \infty$ we obtain that for $p>0$, $\varphi_{\mathrm{max}}\to 0$
and for $p<0$, $\varphi_{\mathrm{max}}\to \infty$. So we conclude that only for $p<0$ the singularity
is a conical singularity (in the usual sense that we get an angular deficit), while
for $p>0$ what we obtain as $\tilde{r}\to 0$ is not a deficit, but instead a decompactification of the angular variable.

Then we have the following cases
\be
\hspace{-15mm}
\ba{rlcll}
\hat{\epsilon}=+1&\displaystyle x\in\left]0,\frac{1}{2}\right[&\Rightarrow&\displaystyle p\in\left]0,\frac{1}{2}\right[:&\mathrm{decompactification\ singularity}\\[5mm]
\hat{\epsilon}=-1&\displaystyle x\in\left]0,\frac{1}{6}\right[&\Rightarrow&\displaystyle p\in\left]0,+\infty\right[:&\mathrm{decompactification\ singularity}\\[5mm]
\hat{\epsilon}=-1&\displaystyle x\in\left]\frac{1}{6},\frac{1}{2}\right[&\Rightarrow&\displaystyle p\in\left]-\infty,-1\right[:&\mathrm{conical\ singularity}
\ea
\ee

\subsection{Trivial Scalar Field Solutions}

The contraction of the Ricci tensor is a constant
\be
R_{\mu\nu}R^{\mu\nu}=\frac{K^4}{4C_f^4}\left(8C_f^4-8C_A^2C_f^2C_h^2+3C_A^4C_h^4\right)
\ee
which indicates that the space-time has no singularities.
Specifically the curvature is
\be
R=-\frac{K^2}{2}\left(4-\frac{C_A^2C_h^2}{C_f^2}\right)=-\frac{m^2}{2}+2\Lambda
\lb{R_trivial}
\ee
and can have either positive or negative values. Taking in account the bounds for the cosmological constant~(\ref{L.ttbound})
we obtain that
\be
\ba{rclrl}
R&<&0&\ \ \ \ \mathrm{for}&\displaystyle\hat{\epsilon}=+1\ \ \mathrm{and}\ \ \Lambda<-\frac{m^2}{2}\\[5mm]
R&<&0&\ \ \ \ \mathrm{for}&\displaystyle\hat{\epsilon}=-1\ \ \mathrm{and}\ \ \Lambda\in\left]-\frac{m^2}{2},\frac{m^2}{4}\right[\\[5mm]
R&=&0&\ \ \ \ \mathrm{for}&\displaystyle\hat{\epsilon}=-1\ \ \mathrm{and}\ \ \Lambda=\frac{m^2}{4}\\[5mm]
R&>&0&\ \ \ \ \mathrm{for}&\displaystyle\hat{\epsilon}=-1\ \ \mathrm{and}\ \ \Lambda\in\left]\frac{m^2}{4},+\infty\right[\ .
\ea
\ee

Therefore we conclude we are in the presence of an extended (non localized) configuration,
there is no singularity, hence this solution cannot be considered as a classical particle.
We recall that for $\hat{epsilon}=+1$ and $\Lambda\geq-m^2/2$ and for $\hat{epsilon}=-1$
and $\Lambda\leq-m^2/2$ there are no allowed solutions.

\setcounter{equation}{0}
\section{Mass, Charge and Angular Momentum\lb{sec.MJQ.f}}

In this section we compute the mass, charge and angular momentum.

\subsection{Non-Trivial Scalar Field Solutions}

As expected the Hamiltonian Constraint ${\mathcal{H}}=0$, Momentum Constraint ${\mathcal{H}}^\varphi=0$ and
Gauss Constraint ${\mathcal{G}}=0$ are obeyed, this is actually a way to check that our calculations are correct.

Using~(\ref{MJQ}) we obtain that the Mass of the configuration is
\be
\ba{rcl}
M&=&\displaystyle\left.\left(2h'+4\,\lambda\,h\,\phi\,\phi'\right)\right|^{r\to\infty}_{r\to \delta_M}=\left.\frac{b^2+16\lambda}{b^2}\,C_h\,\frac{\ln(C_\phi\,r)}{\sqrt{r}}\right|^{r\to\infty}_{r\to \delta_M}=\\[5mm]
 &=&\displaystyle -2\,C_h\,p\,\frac{\ln(C_\phi\,\delta_M)}{\sqrt{\delta_M}}
\ea
\label{ME}
\ee
We introduced a cut-off $\delta_M\ll 1$ because this quantity has a infrared divergence
as we compute the limit of $\delta_M\to 0$.

The charge of this configuration is computed to be
\be
Q_{e}=-\frac{2\,C_{h}\,C_\phi\,C_{E}}{C_{f}}
\lb{QE}
\ee

The constant $C_f$ can be set to unity by a proper redefinition of time $t\to t/C_f$
and the redefinitions of the remaining constants $C_h\to C_h/C_f$ and $\theta\to\theta/C_f$.
So without any loss of generality we set $C_f=1$. However we must remember that
$C_E$ as given in~(\ref{constantsE}) has no defined sign and we must demand that
the electric field has the correct sign when compared with the charge.
From~(\ref{QE}) we conclude that in order $Q_e$ and $C_E$ to have the same sign
we are left only with the possibility of $C_h<0$, then $C_h=-4/3$. Then we rewrite the charge as
\be
Q_e=\pm\frac{2\sqrt{2}}{3}\frac{\sqrt{m^2p-2\Lambda(2-3p)}}{\left((m^2-6\Lambda)(1-3p)\right)^\frac{1}{4}}
\ee
where the $\pm$ accounts for positive and negative charge configurations. $C_E$ must account
for this and the sign is set accordingly
\be
C_E\sim\sign(Q_e)
\ee

As we can see from~(\ref{ME}) this choice of signal for $C_h$ affects the mass sign,
the mass is positive or negative depending on the sign of $p$.
We note that the logarithm in~(\ref{ME}) is negative and therefore the mass is positive when $p<0$
and negative when $p>0$. For $\hat{\epsilon}=+1$ it is always negative, while for
$\hat{\epsilon}=-1$ it is negative for $\Lambda\in]0,m^2/6[$ and positive for $\Lambda\in]m^2/6,m^2/2[$

There is also one interesting point concerning the discrete symmetries time-inversion $T$ and $P$.
Inverting time accounts for choosing $C_f=-1$ such that $t\to-t$. The visible direct effects of
the transformation $C_f\to -C_f$ for our solutions is to invert the sign of $C_A$ and $C_E$
(assuming we have fixed the $\pm$ of $C_E$, see~(\ref{constantsE})). Doing so we revert the
sign of the charge definition as it depends explicitly on $C_f$ as well, see~(\ref{QE})), and although
$C_E\to -C_E$, our charge maintains its sign. Then we have two problems, first the charge and the electric field have now
the wrong relative sign (we are considering $C_h<0$ fixed) and secondly the charge
is not transforming properly under $T$ (see for instance equation~(50) of~\cite{KC}, see also~\cite{dunne}).
Therefore we are forced to transform $C_h\to -C_h$ as well obtaining $C_h>0$.
As a consequence $C_A$ does not actually changes sign (because the ratio $C_h/C_f$ does not
change), this accounts fot $T$ violation due to the Chern-Simons term.
Also we note that by choosing $C_f=-1$ (or transforming $C_f\to-C_f$ and $C_h\to -C_h$)
inverts the mass sign. This is actually expected, we recall
the reader that classically a positron looks like an electron travelling backwards in time. 
As for parity $P$, will account for the transformation $C_h\to -C_h$ which from the above
discussion implies as well $C_f\to -C_f$ and we obtain the same effects.

The Angular Momentum of this configuration is
\be
J=-\frac{2\,C_h^3\,C_A\,(p-1)}{C_f}-J_0=\frac{28\,m}{9}\sqrt{\frac{1-3p}{m^2-6\Lambda}}-J_0
\lb{J}
\ee
where $J_0$ is the background angular momentum and will be computed later.
the sign of $J$ does not depend in the particular configuration, but only
on the relative sign between the Maxwell term ($F^2$) and the Chern-Simons term ($A\wedge F$) as
explained on subsection~\ref{sec.metric}. This means it will change if we consider the transformations
$m\to-m$ and vanishes for $m=0$ (as will be shown it does not vanishes in the limits $m\to 0^\pm$, only for $m=0$).
This is clearly also an effect of $T$ and $P$ violation which is expected when a Chern-Simons
term is present.

So as we have just seen our solutions violate both $T$ and $P$ as expected when a Chern-Simons
term is present. This is explicit on the fact that the signs of $C_A$ and $J$ only depend
on the relative sign between the Maxwell and the Chern-Simons term.

We already computed the angle deficit in the last section such that for $C_h=-4/3$ our background has the
correct angular variable $\varphi\in[0,2\pi[$. Here we still have to compute $J_0$, so we are properly explaining
what are the limits of our solutions when we take the Chern-Simons coefficient to zero, $m\to 0$. 
From the constraint interval we have that it corresponds to $\Lambda\to 0$ (equivalent
to $x=\Lambda/m^2\to 0$ and $p\to 0$). We will analyse this limit
from the definitions~(\ref{constantsE}).
In this limit we obtain from~(\ref{constantsE}) that $C_E\to 0$ therefore
we have necessarily $Q_e\to 0$, also we obtain $C_\phi\to 0$ and $C_A\to -\sign(m)C_f/C_h$.
We note that for $C_A$ the limits on the right and left ($m^{\pm}$) are finite with opposite signs such
that for $x=p=0$ we obtain $C_A=0$. Nevertheless the asymptotic limit are defined only
from the left and from the right such that for the limiting cases $C_A\neq 0$. One obtains from~(\ref{J})
that $J\to -2\sign(m)C_h^2-J_0$. The first term corresponds to the background angular momentum, therefore
we obtain
\be
J_0=-2\sign(m)C_h^2\ .
\ee
As already expected its sign depends on the relative sign
between the Maxwell and the Chern-Simons term and accounts for parity violation.
One obtains by a direct computation that $M\to 0$ and also that the curvature vanishes
everywhere, $R\to 0$. Therefore as background for our configurations we obtain a stationary
rotating flat space without any angle deficit as already studied in the last section. The
background metric is
\be
ds^2=-\frac{1}{r}\,dt^2+dr^2+C_h^2 r\,\left(d\varphi+\left(-\frac{\sign(m)}{C_h}\frac{1}{r}+\theta\right) dt\right)^2\ .
\lb{gback}
\ee

\subsection{Trivial Scalar Field Solutions}

We will now compute the charge, mass and angular momentum for the trivial solution~(\ref{tsolsE})
with $\phi=0$.

The mass of the configuration is null, the charge is
\be
Q_e=-\frac{C_E\,C_h}{C_f}=\pm \frac{C_h}{2}\sqrt{\left|\frac{m^2}{2}+\Lambda\right|}
\lb{tQ}
\ee
and the angular momentum is
\be
J=-\frac{C_A\,C_h^3\,K}{C_f}-J_0=-m\,C_h^2-J_0\ .
\lb{tJ}
\ee
We note that again the sign of the angular momentum only depends on the relative sign
between the Maxwell and Chern-Simons term.

We can solve~(\ref{tQ}) for $C_h$ obtaining
\be
C_h=\frac{2Q_e}{\sqrt{\left|\frac{m^2}{2}+\Lambda\right|}}
\ee
and
\be
J=-\frac{4m\,Q_e^2}{\left|\frac{m^2}{2}+\Lambda\right|}-J_0
\ee
Now the $\pm$ in $C_E$ must be chosen accordingly to the sign of the charge such that we obtain
\be
C_E=\frac{\sign(Q_e)C_f}{2}\sqrt{\left|\frac{m^2}{2}+\Lambda\right|}
\ee
Again we can redefine $t\to t/C_f$ that corresponds to set $C_f=1$.

By computing the limit $m\to 0$ we obtain that $C_A\to 0$, therefore both the charge and angular momentum vanish
and we obtain the flat space
\be
ds^2=-C_fdt^2+dr^2+C_h\,(d\varphi+\theta\,dt)^2\ .
\ee
Using the same procedure we obtain that the angular variable is in the range $\varphi\in[0,1/r^2[$, so this
space has some pathologies.

\section{Summary and Discussion of Results\lb{sec.conc}}

\subsection{Summary of Non-Trivial Solutions}

We will briefly resume the results obtained in this paper. Although we are repeating some of the equations
of the article we think it is necessary in order to assemble and clarify all the results obtained.

We found a electric point particle that can constitute either a naked or dressed
singularity, depending on the parameter choices. The results are presented in terms of $x=\Lambda/m^2$,
the cosmological constant to topological mass squared (Chern-Simons coefficient squared) ratio
and the charge $Q_e$ of the configuration.

The metric, scalar field and gauge field solutions for such configuration are
\bea
\ba{rcl}
ds^2&=&\displaystyle\left(\frac{16}{9}\,r\,\left(C_A\,r^{p-1}+\theta\right)^2-r^{2p-1}\right)\,dt^2+dr^2\\[4mm]
    &+&\displaystyle\frac{16}{9}\,r\,d\varphi^2+\frac{16}{9}\,r\,\left(C_A\,r^{p-1}+\theta\right)\,dt\,d\varphi\\[6mm]
\phi&=&\displaystyle-\frac{2}{b}\ln(|m|\sqrt{\frac{1-6x}{1-3p}}r)\\[6mm]
A_0&=&\displaystyle\frac{C_E}{p-1}r^{p-1}
\ea
\nonumber
\eea
where $\theta$ and $b$ are free parameters and all the remaining constants depend only on the
cosmological constant to Chern-Simons square coefficient ratio $x=\Lambda/m^2$
\bea
\ba{rcl}
p&=&\displaystyle-\frac{3x-\hat{\epsilon}\sqrt{x(2-3x)}}{1-6x}\\[6mm]
C_A&=&\displaystyle \frac{3\sign(m)}{4(1-p)}\,\sqrt{\frac{(1-3p)}{(1-6x)}}\\[6mm]
C_E&=&\displaystyle\frac{\sign(Q_e)\sqrt{p+2x(3p-2)}}{\sqrt{2|m|}}\left(\frac{1-3p}{(1-6x)^3}\right)^\frac{1}{4}
\ea
\nonumber
\eea
The Brans-Dicke coefficient is determined up to the free parameter $b$ as
$\lambda=-8p/b^2$ and the remaining scalar field exponential coefficients are fixed,
$a=0$ and $c=-b/2$.

The charge, angular momentum and mass are
\be
\ba{rcl}
Q_e&=&\displaystyle\pm\frac{2\sqrt{2}}{3}\frac{\sqrt{p-2x(2-3p)}}{\left((1-6x)(1-3p)\right)^\frac{1}{4}}\\[6mm]
J&=&\displaystyle-\frac{28\,\sign(m)}{9}\left(\sqrt{\frac{1-3p}{1-6x}}-1\right)\\[6mm]
M&=&\displaystyle\displaystyle \frac{8p}{3}\,\frac{\ln(C_\phi\,\delta_M)}{\sqrt{\delta_M}}\frac{\ln\left(|m|\sqrt{\frac{1-6x}{1-3p}}\delta_M\right)}{\sqrt{\delta_M}}
\ea
\nonumber
\ee
The mass is infrared divergent and we consider a cut-off proportional to the Planck Length, $\delta_M\sim l_p=\sqrt{G}$,
being $G$ the Newton gravitational constant in natural units.

The curvature is
\be
R=\frac{3\hat{\epsilon}\sqrt{x(2-3x)}+x(-11+48x-12\hat{\epsilon}\sqrt{x(2-3x)})}{2(1-6x)^2\, r^2}
\ee
and there is always a singularity at $r=0$ that
we classify as decompactification or conical singularity depending if
the range of $\varphi$ goes to $\infty$ or $0$ (respectively) in the limit $r\to 0$.

In the table below we present the possible ranges for $x$, $p$, $\Lambda$, the sign of $M$
and the singularity classification. The $\hat{\epsilon}$ refers to the relative sign between
the gauge sector and the gravitational sector.
\bea
\hspace{-12mm}\ba{ccccccc}\hline\\[-3mm]
\hat{\epsilon}&x&p&\Lambda&M&\mathrm{singularity} \\[1mm]\hline
\\[-2mm]
\ \ +1(ghosts)\ \ &\ ]0,1/2[\ &\ ]0,1/2[\ &\ ]0,m^2/2[\ &\ <0\ &decomp.\\[6mm]
-1(standard)&\ ]0,1/6[\ &\ ]-\infty,0[\ &\ ]0,m^2/6[\ &\ <0\ &decomp.\\[5mm]
            &\ ]1/6,1/2[\ &\ ]1,+\infty[\ &\ ]m^2/6,m^2/2[\ &\ >0\ &conical\\\hline
\ea
\nonumber
\eea

And to finalise we list the curvature sign and the existence or not of an horizon at $r=0$
\bea
\hspace{+5mm}\ba{ccccc}\hline\\[-3mm]
\hat{\epsilon}&x&R&M&\mathrm{horizon} \\[1mm]\hline
\\[-2mm]
\ \ +1(ghosts)\ \ &\ ]0,1/2[\ &<0&\ <0\ &no\\[6mm]
-1(standard)      &\ ]0,1/6[\ &<0&\ <0\ &no\\[5mm]
               &\ ]1/6,9/39[\ &<0&\ >0\ &yes\\[5mm]
              &\ ]9/39,9/26[\ &>0&\ >0\ &yes\\[5mm]
               &\ ]9/26,1/2[\ &>0&\ >0\ &no\\\hline
\ea
\nonumber
\eea

So we conclude that there is an horizon at $r=0$ only for standard fields and the $x$ range
\bea
x\in\left]\frac{1}{6},\frac{9}{26}\right[\nonumber
\eea
such that we obtain a dressed singularity. We note that
the mass of the solution are positive in this range.
All remaining cases hold a naked singularity.

\subsection{Summary of Trivial Solutions}

We will summarize only the results for null scalar field ($\phi=0$), i.e. solutions
without the scalar field at all. This case have been addressed in~\cite{Kogan1}
without cosmological constant, we think is worthwhile to review these results with
a non-null cosmological constant.

So, for $\phi=0$ we found a electric extended configuration without singularities.
The results are presented in terms of $K=\sqrt{m^2/2-\Lambda}$ and
the charge $Q_e$ of the configuration.
The metric and gauge field solutions for such configuration are
\bea
\ba{rl}
ds^2&=\left(-\cosh^2(K\,r)+C_h^2(C_A\,\sinh(K\,r)+\theta )^2\right)\,dt^2+dr^2+C_h^2\,d\varphi^2\\[4mm]
    &+2C_h^2\left(C_A\,\sinh(K\,r)+\theta\right)\,dt\,d\varphi\\[6mm]
A_0&=\displaystyle\frac{C_E}{K}\sinh(K\,r)
\ea
\nonumber
\eea
with
\bea
\ba{rcl}
K&=&\displaystyle\sqrt{\frac{m^2}{2}-\Lambda}\\[7mm]
C_h&=&\displaystyle\frac{2Q_e}{\sqrt{\left|\frac{m^2}{2}+\Lambda\right|}}\\[7mm]
C_A&=&\displaystyle\frac{m}{2Q_e\,K}\sqrt{\left|\frac{m^2}{2}+\Lambda\right|}\\[7mm]
C_E&=&\displaystyle\frac{\sign(Q_e)}{2}\sqrt{\left|\frac{m^2}{2}+\Lambda\right|}
\ea
\nonumber
\eea

$K$ can be both real and imaginary. In the case of imaginary $K$ we obtain periodic solutions with period $2\pi/|K|$.
We note that $C_A$ is multiplying by $\sinh(K\,r)$ and correctly is also a pure imaginary such that $g_{t\varphi}$ is real.

The mass of these configurations is null and the angular momentum is
\bea
J=-\frac{4m\,Q_e^2}{\left|\frac{m^2}{2}+\Lambda\right|}\ .
\nonumber
\eea
There are no singularities and the curvature is constant
\bea
R=-K^2+\Lambda=-\frac{m^2}{2}+2\Lambda\ .
\nonumber
\eea

In the next table we list the ranges for $K$, $\Lambda$ and the sign of the curvature
\bea
\hspace{+7mm}\ba{cccc}\hline\\[-3mm]
\hat{\epsilon}&\Lambda&K&R\\[1mm]\hline
\\[-2mm]
\ \ +1(ghosts)\ \ &\ ]m^2,+\infty[\ &\ ]-\infty,-m^2/2[\ &\ <0\\[6mm]
-1(standard)      &\ ]0,m^2[\ &]-m^2/2,0[&\ <0\\[5mm]
               &\ ]0,+\infty[i\ &]0,m^2/4]&\ \leq 0\\[5mm]
               &\ ]9/26,1/2[\ &]m^2/4,+\infty[&\ >0\\\hline
\ea
\nonumber
\eea

\subsection{Discussion of Results}

Given the Einstein Maxwell Chern-Simons theory coupled to a massless gravitational scalar field
with action~(\ref{S}) discussed in section~\ref{sec.EOM} we obtained the above classical solutions with electric charge
only. We study both non-trivial and trivial solutions for the scalar field. For non-trivial
solutions of the scalar field we obtain a rotating electric point particle that
for the opposite sign between the gravitational and gauge sector and a certain range
of the ratio $\Lambda/m^2$ is dressed, while for trivial solutions of the scalar field
we find an extend charge configuration that cannot be interpreted as a particle.

For non-trivial solutions it turns out that the solutions are highly constraint depending on
the cosmological constant to Chern-Simons coefficient squared $x=\Lambda/m^2$ which is constraint
to the range $x\in]0,1/2[$. Further
requiring that the background obtained (in the limit $x\to 0$) to have no angular deficit we obtain
only two free parameters, $\theta$ that accounts for
the globally rotation of space and the $\phi$ exponential coefficient $b$.
Both of them are not relevant for any physical observables.
We study both non-trivial and trivial solutions
for the scalar field. Also we consider both the cases for the relative sign between the gauge sector
and the gravitational sector $\hat{\epsilon}=\pm 1$. When they have the same sign ($\hat{\epsilon}=+1$) we
have that the gauge fields are ghosts in the sense that contribute a negative amount of energy to the Hamiltonian,
while in the case that they have opposite sign ($\hat{\epsilon}=-1$) we have
the standard case. Although the expressions for the solutions are expressed in the same way,
the constants and consequently the physics change significantly. In particular the space-time curvature as well
as the existence or non-existence of horizons will be sensitive to it. 

For trivial solutions, the solutions are given in terms of $K=\sqrt{m^2/2-\Lambda}$ and the charge $Q_e$
and $\theta$ are free parameters. Although the cosmological constant is still bounded by the topological mass these
bounds are not so restrictive. Again the relative sign $\hat{\epsilon}=\pm 1$ between the gravitational
and gauge sector is relevant. In the limit $m\to 0$ we obtain for $\hat{\epsilon}=+1$ that $\Lambda<0$ while
for $\hat{\epsilon}=-1$ that $\Lambda>0$. Our background is flat but with an angular deficit.
In a similar way the curvature is sensitive to the relative sign $\hat{\epsilon}$.

The inclusion of the Chern-Simons topological term introduces very
interesting features. Besides imposing the space to be rotating as explained in section~\ref{sec.metric}
it imposes bounds on the cosmological constant trough the topological mass $m$.
For the non-trivial solutions it constraints the allowed value for the cosmological constant
to the interval $\Lambda\in]0,m^2/2[$ such that the limit $m\to 0$
corresponds also to $\Lambda\to 0$ (equivalent to $x\to 0$ and $p\to 0$)
from the constraint $0<\Lambda<m^2/2$ and we obtain in this limit a flat stationary background space-time.
Then the cosmological constant is turn on and off by the Chern-Simons coefficient.
It is very interesting that these facts emerges only as a consequence of the Chern-Simons
term with out any ha-doc assumption. In this framework the existence of the cosmological
constant can be interpreted as being due to the existence of the scalar field and the
topological massive matter that constitute the electric point-particle.
Therefore we can interpret that the charged matter deforms space-time such that the
deformation is parameterized by the charge $Q_e$ and Brans-Dicke coefficient $\lambda$ and
that the parameter $x$ is given as a function of $Q_e$ and $\lambda$.
As expected this matter affects the curvature, either positively or negatively,
depending on the sign of the gauge sector. For the trivial solution the cosmological
constant bounds are not so restrictive but still exists a relation between topological
mass and cosmological constant bounds, for $\hat{\epsilon}=+1$ we have
$\Lambda<-m^2/2$ and for $\hat{\epsilon}=-1$ we have $\Lambda>-m^2/2$.

As already mentioned, for non-trivial solutions, we have that the cosmological constant is always positive.
However concerning the curvature we have different behaviours depending on the relative sign between the
gauge and gravitational sector. For $\hat{\epsilon}=+1$ the curvature is always positive while for
$\hat{\epsilon}=-1$ the curvature is positive only for high values of $x=\Lambda/m^2\in]9/38,1/2[$.
To understand why let us contract the Einstein equations with the metric such that we obtain
the relation
\bea
R=3e^{b\phi}\Lambda-\lambda(\partial\phi^2)+\hat{\epsilon}e^{c\phi}E^2\ .
\nonumber
\eea
For solutions with $\hat{\epsilon}=+1$ the Brans-Dicke coefficient is always negative,
hence all terms contribute positively to the curvature. For solutions with $\hat{\epsilon}=+1$
we have that the electric field contribution is always negative and that the Brans-Dicke coefficient is positive
when $x\in]0,1/6[$ and negative when $x\in]1/6,1/2[$. Therefore we have the following cases, for $x\in]0,1/6[$ both the scalar
field and electric field contribute negatively
to the curvature while for $x\in]1/6,1/2[$ the scalar field contributes positively and the electric field contributes
negatively to the curvature. We further note that from the expressions for the several
constants~(\ref{constantsE}) for the allowed solutions~(\ref{solsE}), we have that near $x=1/6$ the electric
field contribution is predominant when compared with the scalar field contribution (that is negletable, $C_\phi\to 0$).
So only away from $x=1/6$ the scalar matter will become dominant over the charged matter and we have a positive
curvature for $x\in]9/38,1/2[$.
In this way we conclude that the scalar field is determinant in imposing the bounds on the cosmological constant (on the
non-trivial solutions). Also it is the scalar field that allows for the existence of horizons. We concluded that there
are horizons only for $\hat{\epsilon}=-1$ in the range $1/6<x<9/26$ which corresponds to the greater
positive values of $p$ ($>3/2$), remembering that the Brans-Dicke coefficient is proportional to $\lambda\sim p$
this means that these values correspond to a region in which the scalar matter contributes positively to the
curvature.

For the trivial solutions although the bound on the cosmological constant
is not so restrictive the same behaviour concerning the curvature applies as can be
seen directly in the expression for the curvature~(\ref{R_trivial}) that depends both in the cosmological constant
and topological mass. For the trivial solutions we will have positive curvature only for
$\hat{\epsilon}=-1$ and $\Lambda>m^2/2$.

The charges and angular momenta of the configurations are finite. The solutions are, for both non-trivial and
trivial solutions of the scalar field, rotating spaces with angular momenta $J\sim m$ (or $J\sim\sign(m)$), this accounts explicitly
for the known parity $P$ and time-inversion $T$ violation due to the Chern-Simons term~\cite{qed3_02}.
Is explicit in the sense that the sign of the constant $C_A$ and of the angular momentum
only depends on the relative sign between the Chern-Simons coefficient and the gravitational
curvature term.

Concerning the mass of our configurations we concluded that its positiveness (or negativeness) is sensitive to
the relative sign between the gravitational and gauge sector.
However these results are not conclusive, although the charge and angular momentum are finite,
the mass is infrared divergent, this is the main drawback of our solutions. The background is
flat and therefore the reference mass (of the background) is null. Here we consider a cut-off
of the order of the Planck Length. We believe that something is still missing in our theory,
as already explained previously we are not considering a gravitational Chern-Simons.
This correction to the Einstein action induces a correction to the configuration mass and
would regularize it~\cite{CS_grav_03,CS_grav_04,CS_grav_05,CS_grav_06,CS_grav_07,CS_grav_08}.
For the extended trivial solutions the mass is null.

As a final remark we note that our solutions hold that $a=0$. Therefore the gravitational
sector resembles an action with a dilatonic potential
given by our cosmological constant term in~(\ref{S}), see for instance~\cite{CM_01,CM_02,dilpot}.
We notice that by setting $a=0$ the field $\phi$ is only
minimally coupled to the $2+1$ metric and all the fields are expressed in terms of the scalar field
(see the derivation of the solutions in section~\ref{sec.E}), therefore we would expect to obtain
similar results by including more generic dilatonic potentials. An important point to stress here
is that although our action is similar to the  action of the work of Chan and Mann~\cite{CM_01,CM_02}~(CM)
with an extra Chern-Simons term, it is not possible to obtain the solutions
of those works in the limit $m\to 0$. The main reason is that the although there the CM action is generic the authors only consider solutions
for the particular case in which the scalar field can be interpreted as a dilaton.
This means that the constants in our action~(\ref{S}) would be $a=0$, $c=-b=4$ and $\lambda=8$ which is not the case since our constants are related as
$c=-b/2$ and $\lambda$ is dependent on several parameters. Therefore our massless scalar field cannot be interpreted as a dilaton.
Secondly in our case we have no horizons away from $r=0$ and both our cosmological constant $\Lambda$ and charge $Q_e$ vanishes in the
limit $m\to 0$ as already explained in detail, therefore we cannot possible obtain the solutions of Chan and Mann
since their horizons are set uniquely by $\Lambda$ and $Q_e$.

Interesting enough our gravitational field $\phi$ can be related to the works of polarized cylindrical
gravitational waves in $3+1$ gravity~\cite{QG_01,QG_02,QG_03,QG_04}. For an explicit form of the effective
$2+1$ dimensional action see for instance equation~(1) of~\cite{QG_04} (see also~\cite{QG_03}). In our case
we further have a full gauge sector such that our classical solutions could constitute a possible electric
charged background with cilindrical symmetry in $3+1$ dimensions (our solutions would correspond then to a
electric charged string). Also similar actions have been considered in cosmological scenarios~\cite{BB}
and in brane worlds~\cite{branes_01}.

After finishing this work the author realized that after we get our solutions redefining
the radial coordinate accounts for changing the dilaton coupling (for $a\neq 0$) with the curvature $R$
and the Brans-Dicke parameter, however they will have generally different exponential factors,
this does not invalidate the work presented here, simply we could yet consider a more generic action.

As an extension to this work the author intends to compute a pure magnetic solution~\cite{csII}
using a similar action and procedure to this article.
In order such configuration to exist it is necessary to consider an external electric charge
distribution because as can be seen explicitly from Maxwell equations~(\ref{EOM}) or~(\ref{E.M2})
for $E=E_*=0$ we have that $B\sim j^0$. If we set $j^0=0$ the equations of motion
hold that the magnetic field is null. This discussion is already put forward by Kogan~\cite{Kogan1}
(see conclusions of this reference). Another possible way out is to consider $Ef=hAB$ (such that $E_*=0$, see~(\ref{E.EB*})
in appendix). In these cases the rotation will induce a electric field (see discussion in section~\ref{sec.MJQ}).
Also as other possible direction of research it would be
interesting to consider extensions of this work that include gravitational
Chern-Simons (as already explained we would expect to obtain finite mass)
and dilatonic potentials.\vfill

\vspace{5mm}\noindent {\large\bf Acknowledgments}\\
This work was supported by Grant SFRH/BPD/5638/2001, SFRH/BPD/17683/2004 and POCTI/P-FIS-57547/2004.
The author thanks both Ian Kogan for pointing out~\cite{Kogan1} and Bayram Tekin for suggesting
to further investigate gravitational solutions of Chern-Simons theories.
The author thanks Jos\'e Sande Lemos and \'Oscar Dias for invaluable discussions,
also for explaining in detail part of their works~\cite{SLO_01,SLO_02,SLO_03,SLO_04}
as well recommending literature on the subject. The author thanks Nuno Reis and Carlos Herdeiro
for several discussions and suggestions. The author thanks Paulo Vargas Moniz and Dmitri Gal'tsov
for important remarks and suggestions.

\appendix
\setcounter{equation}{0}
\section{Cartan Formalism\lb{A.cartan}}

In this appendix we study the equations of motion in the Cartan Frame.

The Lagrangean 3-form corresponding to the action~(\ref{S}) is rewritten as
\be
\ba{rl}
{\mathcal{L}}=&-\left\{e^{a\phi}\left[R*1+2\lambda\,d\phi\wedge*d\phi\right]-e^{b\phi}\Lambda*1\right.\\[4mm]
              &\left.+\hat{\epsilon}e^{c\phi}\left[F\wedge*F+*J\wedge A\right]+\hat{\epsilon}\frac{m}{2}\,A\wedge F\right\}
\ea
\ee
with $R$ the metric curvature and $F=dA$ and where we define the Hodge dual as usual
\be
(*X)^{i_1\ldots i_q}=(-1)^D\,\frac{\sqrt{-g}}{p!}\epsilon^{i_1\ldots i_q\,j_1\ldots j_p}X_{j_1\ldots j_p}
\ee

Introducing a triad $\{e^0,e^1,e^2\}$ such that
\be
e^i=e^i_{\ \alpha} dx^\alpha\ \ \ \ g_{\alpha\beta}=\eta_{ij}e^i_{\ \alpha}e^j_{\ \beta}
\ee
where the Greek indices refer to the coordinates $(x^0=t,x^1=r,x^2=\varphi)$ and the
roman ones to the Cartan frame triad (meaning the flat space indices).

Varying the Lagrangean with respect to the Gauge field $A$, the coframe field $e^i$ and the
dilaton $\phi$ we obtain the equations of motion in the Cartan frame
\be
\hspace{-15mm}
\ba{rcl}
d(*F\,e^{c\phi})-*J&=&\displaystyle-\frac{m}{2}\,F\\[5mm]
\displaystyle \left[e^{a\phi}\left(G_{ij}+\Phi_{ij}\right)-e^{b\phi}\eta_{ij}\Lambda*e_i-2e^{c\phi}T_{ij}\right]*e^j&=&0\\[5mm]
\displaystyle e^{a\phi}\left[(4a^2-\lambda)d*d\phi+a(4a^2-2\lambda)d\phi\wedge *d\phi\right]& &\\[4mm]
-(b-3a)e^{b\phi}\Lambda*1&=&\displaystyle \hat{\epsilon}2(a+c)e^{c\phi}F\wedge*F
\ea
\ee
respectively the Maxwell, Einstein and scalar field equations. We will specify the Einstein tensor $G_{ij}$, the
Energy-Momentum tensor $F_{ij}$ and the scalar field tensor $\Phi_{ij}$ for each metric parameterization used
\be
\hspace{-15mm}
\ba{rcl}
G_{ij}&=&R_{ij}-\frac{1}{2}\eta_{ij}R\\[5mm]
T_{ij}&=&\displaystyle \hat{\epsilon}\left(F_{ik}F_j^{\ k}-\frac{1}{4}\eta_{ij}F_{kl}F^{kl}\right)\\[5mm]
\Phi_{ij}&=&\displaystyle -a\nabla_i\partial_j\phi+a\,\eta_{ij}\nabla^2\phi+(\lambda-a^2)\partial_i\phi\partial_j\phi-\left(\frac{\lambda}{2}-a^2\right)\eta_{ij}\partial_k\phi\partial^k\phi
\ea
\ee

To proceed further one has to introduce a spin connection $\omega^{ij}_{\ \ \alpha}$ and define the corresponding
connection 1-form
\be
\omega^i_{\ j}=\omega^i_{\ j\alpha}dx^\alpha=\omega^i_{\ jk}e^k
\ee

Using the antisymmetric property (from definition)
\be
\omega^{ij}=-\omega^{ji}
\ee
and the Cartan Structure equation
\be
de^i+\omega^i_{\ j}\wedge e^j=0
\ee
is enough to determine all the connection coefficients $\omega^i_{\ jk}$.
In this work we are considering only radial symmetric configurations and metric
parameterization such that $e^1=dr$ (note that a redefinition
of $r$ introduces a non trivial metric component $g_{11}$) and $e^0$ and $e^2$ depend only on $dt$ and
$d\varphi$ (means that the metric has nonnull components $g_{\alpha\alpha},g_{02}$).
In these particular cases we get the non vanishing connection coefficients
\be
\ba{c}
\omega^{0}_{\ 12}=\omega^{1}_{\ 02}=\omega^{1}_{\ 20}=-\omega^{2}_{\ 10}\ \ \ \ \omega^{0}_{\ 21}=\omega^{2}_{\ 01}\\[5mm]
\omega^{0}_{\ 10}=\omega^{1}_{\ 00}\ \ \ \ \omega^{0}_{\ 20}=\omega^{2}_{\ 00}\\[5mm]
\omega^{0}_{\ 22}=\omega^{2}_{\ 02}\ \ \ \ \omega^{1}_{\ 22}=-\omega^{2}_{\ 12}
\ea
\lb{wcon}
\ee
plus the two equations
\be
\ba{c}
de^0+\omega^0_{\ 1}\wedge e^1+\omega^0_{\ 2}\wedge e^2=0\\[5mm]
de^2+\omega^2_{\ 0}\wedge e^0+\omega^2_{\ 1}\wedge e^1=0
\ea
\lb{Cartaneq}
\ee

Also note that in this case the only Electric field component is $E=F_{01}$ ($F_{02}=0$ from Maxwell equations)
and all the derivatives are with respect to $r$ only. Then it is now possible
to define $T_{ij}$ and $\Phi_{ij}$ for our parameterization:
\be
\ba{rcl}
2T_{00}&=&\hat{\epsilon}\left(B^2+E^2\right)\\[3mm]
2T_{11}&=&\hat{\epsilon}\left(B^2-E^2\right)\\[3mm]
2T_{22}&=&\hat{\epsilon}\left(B^2+E^2\right)\\[3mm]
2T_{02}&=&-2\hat{\epsilon}BE
\ea
\lb{Tij}
\ee
the square of the Maxwell tensor is
\be
F^2=2\hat{\epsilon}(B^2-E^2)
\ee
and
\be
\ba{rcl}
\Phi_{00}&=&-a\phi''+(\lambda/2-a^2)(\phi')^2\\[3mm]
\Phi_{11}&=&\lambda/2(\phi')^2\\[3mm]
\Phi_{22}&=&a\phi''-(\lambda/2-a^2)(\phi')^2
\ea
\lb{Phiij}
\ee

Note that the original electric field $E_{*\,\alpha}=F_{t\alpha}$,
magnetic field $B_*=F_{r\varphi}$ and external current $*J$ are related to the Cartan frame ones $E_i=F_{0i}$, $B=F_{12}$ and $*j$
either by using the triad $e^i_\alpha$ or by the definition of the 2-forms $F=F_{\alpha\beta}\,dx^\alpha\wedge dx^{\beta}=F_{ij}\,e^i\wedge e^j$ and
$*J=\sqrt{-g}\,\epsilon_{\mu\nu\rho}\,J^\mu\,dx^\nu\wedge dx^\rho=\epsilon_{ijk}\,j^i\,e^j\wedge e^k$.

We use the metric parameterization such that the line element is given by
\be
ds^2=-f^2dt^2+dr^2+h^2(d\varphi+Adt)^2
\lb{E.ds}
\ee
such that the usual components read
\be
\ba{rcl}
g_{00}&=&-f^2+h^2A^2\\
g_{11}&=&1\\
g_{22}&=&h^2\\
g_{02}&=&h^2A
\ea
\ee

The Cartan triad is then given by
\be
\ba{rclclll}
\e^0=d\theta^0&=&fdt&\ \ \ \ &e^0_{\ 0}=f&e^0_{\ 1}=0&e^0_{\ 2}=0\\[5mm]
\e^1=d\theta^1&=&dr&\ \ \ \ &e^1_{\ 0}=0&e^1_{\ 1}=1&e^1_{\ 2}=0\\[5mm]
\e^2=d\theta^2&=&h(d\varphi+Adt)&\ \ \ \ &e^2_{\ 0}=hA&e^2_{\ 1}=0&e^2_{\ 2}=h
\ea
\lb{E.e-theta}
\ee
such that the line element is now
\be
ds^2=e^ie_i=\eta_{ij}d\theta ^id\theta ^j=-(d\theta ^0)^2+(d\theta ^1)^2+(d\theta ^2)^2
\lb{E.Cds}
\ee
with Minkowski metric $\eta=\diag(-1,1,1)$.

The original Electric $E_*$ and Magnetic fields $B_*$ are given by
\be
\ba{rcl}
E_*&=&E\,f-B\,h\,A\\[5mm]
B_*&=&B\,h
\ea
\lb{E.EB*}
\ee
where $E$ and $B$ are the Cartan frame Electric and Magnetic fields.

The Cartan external currents $j^i$ are given by
\be
\ba{rclcl}
j^0&=&\displaystyle f\,J^t&=&\displaystyle\frac{e^{-c\phi}}{h}{\mathcal{J}}^t\\[5mm]
j^2&=&\displaystyle h\,\left(J^\varphi-A\,J^t\right)&=&\displaystyle\frac{e^{-c\phi}}{f}\left({\mathcal{J}}^\varphi-A\,{\mathcal{J}}^t\right)
\ea
\lb{j.J}
\ee
where $J^\mu$ are the original external currents.  For
radial currents one has simply $j^1=J^1=e^{-c\phi}{\mathcal{J}}^r/hf$. We note that
in terms of the physical ${\mathcal{J}}^\mu$ (measured by an external observer)
we have $J^\mu=e^{c\phi}{\mathcal{J}}^\mu/hf$ (see eq~(\ref{Jdens})).

From the form differentials
\be
\ba{rcl}
de^0&=&-\alpha e^0\wedge e^1\\[3mm]
de^2&=&\beta e^1\wedge e^2-\gamma e^0\wedge e^1
\ea
\ee
we conclude that, except for the external currents, the Equations of motion, connections, curvature and so on depend only
on the combinations
\be
\alpha=\frac{f'}{f}\ \ \ \ \beta=\frac{h'}{h}\ \ \ \ \ \gamma=\frac{h\,A'}{f}
\lb{E.abgdef}
\ee

We list the non null connections in the Cartan frame
\be
\ba{l}
\omega^0_{\ 10}=\omega^1_{\ 00}=\alpha\\[5mm]
\omega^0_{\ 12}=\omega^0_{\ 21}=\omega^1_{\ 02}=\omega^1_{\ 20}=\omega^2_{\ 01}=-\omega^2_{\ 10}=-\gamma/2\\[5mm]
\omega^1_{\ 22}=-\omega^2_{\ 12}=-\beta\\[5mm]
\ea
\ee
and the Einstein tensor components
\be
\ba{rcl}
G_{00}&=&-\beta^2-\gamma^2/4-\beta'\\[3mm]
G_{11}&=&\alpha\beta+\gamma^2/4\\[3mm]
G_{22}&=&\alpha^2-3\gamma^2/4+\alpha'\\[3mm]
G_{02}&=&-\beta\gamma-\gamma'/2
\ea
\lb{E.Gij}
\ee

Then the Maxwell Equations are
\bea
B'+\alpha B+c\,B\,\phi'-\gamma\,E-j^2&=&\displaystyle-m\,E\,e^{-c\phi}\lb{E.M1}\\[5mm]
E'+\beta E+c\,E\,\phi'+j^0&=&\displaystyle-m\,B\,e^{-c\phi}\lb{E.M2}
\eea

The Einstein Equations are
\bea
\hspace{-19mm}\displaystyle e^{a\phi}\left(\beta\gamma+\frac{\gamma'}{2}\right)&=&\displaystyle 2\hat{\epsilon}e^{c\phi}E\,B\lb{E.E1}\\[5mm]
\hspace{-19mm}\displaystyle e^{a\phi}\left[\beta^2+\frac{\gamma^2}{4}+\beta'+a\phi''+\left(a^2-\frac{\lambda}{2}\right)(\phi')^2\right]+\frac{1}{2}e^{b\phi}\Lambda&=&\displaystyle -\hat{\epsilon}(B^2+E^2)e^{c\phi}\lb{E.E2}\\[5mm]
\hspace{-19mm}\displaystyle e^{a\phi}\left[\alpha^2-\frac{3\gamma^2}{4}+\alpha'+a\phi''+\left(a^2-\frac{\lambda}{2}\right)(\phi')^2\right]+\frac{1}{2}e^{b\phi}\Lambda&=&\displaystyle \hat{\epsilon}(B^2+E^2)e^{c\phi}\lb{E.E3}\\[5mm]
\hspace{-19mm}\displaystyle e^{a\phi}\left[\alpha\beta+\frac{\gamma^2}{4}+\frac{\lambda}{2}(\phi')^2\right]+\frac{1}{2}e^{b\phi}\Lambda&=&\displaystyle \hat{\epsilon}(B^2-E^2)e^{c\phi}\lb{E.E4}
\eea
and the dilaton equation is
\be
\hspace{-19mm}e^{a\phi}\left[(4a^2-\lambda)\phi''+a\left(4a^2-2\lambda\right)(\phi')^2\right]+(3a-b)e^{b\phi}\Lambda=\hat{\epsilon}(a+c)(B^2-E^2)e^{c\phi}
\lb{E.D}
\ee

\end{document}